# The Impact of Uncertainties in Reaction Q-values on Nucleosynthesis in Type I X-Ray Bursts

A. Parikh[1,2], J. José[2,3], C. Iliadis[4,5], F. Moreno[2] and T. Rauscher[6]

[1]Physik Department E12, Technische Universität München, D-85748 Garching, Germany
[2] Departament de Física i Enginyeria Nuclear, EUETIB, Universitat Politècnica de Catalunya, E-08036 Barcelona, Spain
[3]Institut d'Estudis Espacials de Catalunya (IEEC), E-08034 Barcelona, Spain
[4]Department of Physics and Astronomy, University of North Carolina, Chapel Hill, NC 27599-3255, USA
[5]Triangle Universities Nuclear Laboratory, Durham, NC 27708-0308, USA
[6]Department of Physics, University of Basel, CH-4056 Basel, Switzerland

**ABSTRACT**

Nucleosynthesis in Type I X-ray bursts may involve up to several thousand nuclear processes. The majority of these processes have only been determined theoretically due to the lack of sufficient experimental information. Accurate reaction Q-values are essential for reliable theoretical estimates of reaction rates. Those reactions with small Q-values (< 1 MeV) are of particular interest in these environments as they may represent waiting points for a continuous abundance flow toward heavier-mass nuclei. To explore the nature of these waiting points, we have performed a comprehensive series of post-processing calculations which examine the sensitivity of nucleosynthesis in Type I X-ray bursts to uncertainties in reaction Q-values. We discuss and list the relatively few critical masses for which measurements could better constrain the results of our studies. In particular, we stress the importance of measuring the mass of $^{65}$As to obtain an experimental Q-value for the $^{64}$Ge(p,γ)$^{65}$As reaction.

PACS numbers: 21.10.Dr, 26.30.Ca, 26.60.Gj, 97.80.Jp

I.  INTRODUCTION

Soon after the discovery of the phenomena now known as Type I X-ray bursts [1,2], the underlying mechanism was generally established as unstable burning of accreted solar-type material on the surface of a neutron star in a low-mass binary system (e.g., [3 – 6]) . The stellar binary system is close enough to allow mass transfer episodes. This matter flow forms an accretion disk that surrounds the neutron star and ultimately accumulates on its surface, building up an envelope in semi-degenerate conditions. As material piles up on top of the neutron star, the envelope is heated up without any significant expansion due to degeneracy, driving a violent thermonuclear runaway (for reviews see [7 – 9]). To date, 84 Galactic X-ray burst sources have been identified [10].

Type I X-ray bursts (hereafter XRBs) are typically characterized by burst energies of $10^{39} – 10^{40}$ ergs ($L_{peak}$ ~ 3 x $10^{38}$ ergs/s), timescales of 10 – 100 s, and recurrence times of hours – days (see e.g., [11]). The maximum temperatures (T ~ 1 – 2 GK) and densities (ρ ~ $10^6$ g/cm$^3$) thought to be achieved in XRBs eventually drive the nucleosynthesis in these events along the proton-rich side of stability through the αp and rp processes, reaching A > 60 and perhaps even A > 100 (e.g., [12 – 19]). Note that the properties of XRBs are particularly dependent on the accretion rate $\dot{M}_{acc}$(e.g., [6,9]): here, we focus on behaviour resulting from the "intermediate" regime of $\dot{M}_{acc}$ ~ 4 x $10^{-10}$ – 2 x $10^{-8}$ M$_\odot$/yr, where bursts are thought to arise from both hydrogen and helium burning. Weaker flashes (with corresponding reduced nucleosynthetic flow) may be expected for $\dot{M}_{acc}$ < 4 x $10^{-10}$ M$_\odot$/yr, while $\dot{M}_{acc}$ > 4 x $10^{-10}$ M$_\odot$/yr may eventually lead to stable burning on the surface of the neutron star.

Realistic models of these phenomena are challenging because of the extreme astrophysical conditions associated with XRBs (requiring detailed, computationally-intensive hydrodynamic studies), the extent of the nucleosynthesis (requiring complex reaction networks involving several hundred isotopes and several thousand nuclear processes), and the lack of experimental nuclear physics information along the nucleosynthesis path. While several recent investigations [18 – 25] have overcome the simplifications of past studies [13 – 17, 26 – 28] by improving the reaction network and/or underlying astrophysical model employed, uncertainties due to the nuclear physics input persist.

Nuclear physics uncertainties may certainly affect predicted XRB properties, whether directly through the influence of particular rates [18, 20 – 22, 24, 29 – 31], or indirectly through the effects of accretion onto the particular nucleosynthetic products of previous bursts (compositional inertia – see [18, 25, 32]); sedimentation of burst ashes may also affect the ignition of future bursts [33]. XRB nucleosynthesis will affect thermal, electrical and mechanical properties of the neutron star crust [14, 34 – 37], which are important for understanding the evolution of the neutron star magnetic field and the quiescent X-ray binary luminosity between bursts. As well, reliable predictions of carbon production in XRBs are essential for testing carbon-ignition models of superbursts (e.g., [37 – 40]). Finally, although ejection of material during XRBs may be possible during photospheric radius expansion [41, 42], it is unlikely due to the strength of the neutron star gravitational field. Nonetheless, the nature of XRB nucleosynthesis may still be characterized through high-resolution X-ray spectra [41, 43 – 46], for example by searching for redshifted photoionization edges.

Recently, we performed a comprehensive and systematic set of calculations to examine how uncertainties in nuclear physics processes –specifically thermonuclear reaction rates and weak interaction rates – may affect XRB nucleosynthesis [47]. In that work we compiled a list of key reaction rates, measurements of which would help to constrain our predictions of XRB nucleosynthesis. In the present work we now consider the particular effects of uncertainties in reaction Q-values on XRB yields, again in the hope of motivating and focusing experimental efforts.

## II. TECHNIQUE AND FRAMEWORK

Through the rp process, the reaction flow during an XRB (largely (p,γ) reactions) is eventually driven toward the proton drip-line: capture of successive protons by nuclei along the drip-line is inhibited by a strong reverse photodisintegration reaction. The competition between the rate of proton capture and the rate of β-decay at these 'waiting points' determines the extent of the abundance flow to heavier masses during the burst. Explicitly, these waiting points arise because the proton-capture reactions on these nuclides have sufficiently small Q-values (relative to XRB temperatures – at 1 GK, kT ~ 100 keV) that an equilibrium between the rates of the forward (p,γ) and reverse (γ,p) processes is quickly established. According to detailed balance, for a reaction X(p,γ)Y, the rate $N_A\langle\sigma v\rangle$ and the photodisintegration decay constant λ are related by [48]:

$$\frac{\lambda(Y+\gamma \to X+p)}{N_A\langle\sigma v\rangle(X+p \to Y+\gamma)} = 9.8685\times 10^9 T_9^{3/2} \frac{g_X g_p}{g_Y}\left(\frac{G_X G_p}{G_Y}\right)\left(\frac{M_X M_p}{M_Y}\right)^{3/2} e^{-11.605Q/T_9} \quad (1)$$

where $g_i = 2J_i + 1$ and $M_i$ are statistical factors and ground-state masses (in u) for nuclei of ground-state spin $J_i$, Q is the ground-state Q-value of the forward reaction X + p → Y + γ (in MeV), $T_9$ is the temperature in GK and $G_i$ are normalized partition functions (this expression only holds for stellar rates). The Q-value enters exponentially in the above expression and thus is clearly the most important nuclear physics information needed to characterize the rate equilibrium. Consequently, the leakage of material to heavier nuclei (via subsequent proton-capture on the equilibrium abundance of Y) is critically dependent on the Q-value of the reaction X(p,γ)Y when this Q-value is sufficiently small.

In this study we have explored the effects of individually varying each reaction with Q < 1 MeV in our XRB network within its Q-value uncertainty ΔQ. For each case (Q ± ΔQ for a total of 200 reactions) we ran a one-zone post-processing calculation and compared final XRB yields with calculations using standard Q-values. In addition, ten different XRB scenarios were employed (sampling the parameter space of possible XRBs – see Table I) to fully explore any possible impact of each Q-value variation; these scenarios have been described in detail in [47]. Briefly, three of these ten scenarios (K04, F08, S01) use thermodynamic histories and conditions from Koike et al. [17], Fisker et al. [19] and Schatz et al. [15], respectively. The accretion rates assumed in these three models lie in the range $\dot{M}_{acc}$ = 2 x 10$^{-9}$ – 2 x 10$^{-8}$ M$_\odot$/yr; these fall into the intermediate regime of $\dot{M}_{acc}$ as discussed in Section I. Four other scenarios are based upon K04, but scaled to different peak temperatures (hiT, lowT) and burst durations (long, short). (Here, we take 'burst duration' to be the characteristic timescale of the temperature and density vs. time thermodynamic histories.) Finally, three further scenarios used the conditions of K04 but with different initial accreted compositions (lowZ, hiZ, hiZ2). A total of 200 x 2 x 10 = 4000 individual XRB post-processing calculations were therefore performed in the present study.

Our XRB nuclear reaction network has been described in [47] and includes the modifications discussed below. In brief, it comprises 606 nuclides between $^1$H and $^{113}$Xe, together with all charged-particle induced reactions between these species. Beta-decay rates from [49] have been used, and the impact of β-delayed nucleon emission has been considered. Experimental rate determinations are used whenever possible (e.g., [50, 51]), but for the majority of reactions in our network the rates have been calculated using the Hauser-Feshbach formalism [52 – 54] due to the lack of sufficient experimental information. All reaction rates incorporate the effects of thermal excitations in the target nuclei. The majority of the Q-values we have varied have only been estimated from systematic trends in the literature. For all cases which were varied, we have assumed the Q-values and uncertainties ΔQ tabulated in the most recent Atomic Mass Evaluation [55]. We emphasize that the (yet-unmeasured) Q-values from [55] are not based on a theoretical mass model, but are derived from the *extrapolation* of experimental masses.

It is important to note the role of the reaction Q-value: namely, it enters twice, once in the Hauser-Feshbach forward rate calculation and again in the reverse rate calculated according to eq. (1). In our study we used theoretical rates calculated with a modified version of the NON-SMOKER code [52, 54]. This new version includes not only changes in the numerical treatment of transitions but also a number of updated nuclear properties: masses from [55], experimental information on ground- and excited-state spins and parities from [56], an improved prediction of ground-state properties when no experimental information is available, and a parity-dependent level density [57]. For reactions with Q < 1 MeV, additional rate sets were consistently computed for *forward and reverse* rates with the same code by using the upper and lower limits of the Q-values according to [55]. A simpler, approximate method to deal with a Q-value change would have been to preserve the forward rate and only recalculate the reverse rate using the new Q-value. For reactions with Q < 1 MeV and ΔQ ~ few hundred keV, these two methods resulted in only small differences in XRB final abundances. This is to be expected for two reasons. First, for such small Q-values and small ΔQ, the change in the forward rate remains small compared to the exponential effect of ΔQ on the reverse rate. Second, around the peak temperatures reached in our calculations, a (p,γ)-(γ,p) equilibrium is achieved which makes the equilibrium abundances insensitive to individual rates, leaving only a sensitivity to the assumed Q-value. The latter is underscored by the fact that our results are quite insensitive to the choice of rate set as proved by test calculations using theoretical rates calculated with the MOST code [53].

We would like to emphasize that one has to be very careful when varying Q-values in a reaction network calculation. Consider, as an example, the nucleus $^{65}$As. A variation in its mass causes changes in the Q-values of the reactions $^{64}$Ge(p,γ)$^{65}$As, $^{65}$As(γ,p)$^{64}$Ge, $^{65}$As(p,γ)$^{66}$Se and $^{66}$Se(γ,p)$^{65}$As. Thus it becomes clear that these Q-values are correlated. However, in our procedure we vary the Q-value for each forward reaction at a time. Although we carefully took the Q-value correlation between forward and corresponding reverse reactions into account (that is, between $^{64}$Ge(p,γ)$^{65}$As and $^{65}$As(γ,p)$^{64}$Ge in the above example), we disregarded the Q-value correlation between subsequent reactions ($^{64}$Ge(p,γ)$^{65}$As and $^{65}$As(p,γ)$^{66}$Se, in the above example). Fortunately, these correlations can be studied analytically [48]. We explored these effects in the above example, namely, we looked at the effect on the $^{64}$Ge decay constant from a) consistently changing the $^{65}$As mass (according to limits in [55]) for both the $^{64}$Ge(p,γ) and $^{65}$As(p,γ) reactions vs. b) just changing the $^{65}$As mass for the $^{64}$Ge(p,γ) reaction and leaving the $^{65}$As mass unchanged in the $^{65}$As(p,γ) reaction. We find that neglecting the change in the Q-value of the $^{65}$As(p,γ) reaction (method (b), corresponding to the approach in the present study) leads to a change of up to 40% in the $^{64}$Ge decay constant relative to the 'consistent' approach (i.e., method (a)), for the temperatures achieved in our models. Similar results were seen when we analytically examined the effect on the $^{68}$Se decay constant due to neglecting the Q-value correlation between the $^{68}$Se(p,γ)$^{69}$Br and $^{69}$Br(p,γ)$^{70}$Kr reactions via the uncertainty in the $^{69}$Br mass. Since the goal of this work was to identify the main reactions whose Q-values are sufficiently uncertain so as to significantly impact final XRB yields (i.e., by at least a factor of two, see section III), our approach should be adequate. This was confirmed through test post-processing calculations that examined the effects of these Q-value correlations (i.e., method (a) vs. method (b)) in more detail, and found our results (i.e., Tables III and IV, see section III) to be robust.

III.    RESULTS AND DISCUSSION

Table II lists the final abundances (as mass fractions $X_{f,std}$) calculated with each of the ten XRB models described in Section II, using the *standard* reaction rates in our network (i.e., before varying any reaction Q-values). Only stable isotopes or those with $t_{1/2} > 1$ hr are given (the rest are assumed to fully decay at the end of the burst, and are consequently added to the nearest stable or long-lived daughter nuclei). In addition, these mass fractions $X_{f,std}$ are plotted against mass number A in Figs. 1 – 4 for each of the ten XRB models. If we consider only nuclei with a relatively large post-burst abundance (mass fraction $X_f > 0.01$, see Table I), we find that the nuclear activity reaches its greatest extent in the 'S01', 'hiT' and 'long' models (up to the Ag-Cd region). Indeed, use of a larger nuclear network than the one we have adopted may show additional limited yields ($X_f < 10^{-5}$) at masses above A=113 for these three models, particularly 'S01'. This is a direct result of the higher peak temperatures and/or longer exposure times to high temperatures in these models. For these reasons, as well as that of hydrogen exhaustion, the nuclear activity is more subdued in models 'F08', 'short', 'hiZ' and 'hiZ2' (reaching the Ge-Se region). We also find that the final yield distributions from the 'K04' and 'lowZ' models are quite similar, suggesting a lack of sensitivity to initial metallicity below a certain threshold. Note that for the models adopted directly from the literature (K04, F08, S01) the most abundant isotopes from our calculations (see Tables I and II) are in good qualitative agreement with the results from the respective articles [17,19,15] for the most part (but see also [47]).

We stress that the abundances of Table II and Figs. 1 – 4 arise from post-processing calculations (where the reaction network is not coupled to hydrodynamics, and hence, convection is neglected). For this reason, these absolute abundances are generally useful only for the purpose of comparison with other one-zone post-processing calculations. More complex treatments (such as fully self-consistent hydrodynamic calculations) are required for reliable absolute final XRB abundances. The rough agreement between abundances from our post-processing work and those from the hydrodynamic and hybrid models of 'F08' and 'K04', respectively, is encouraging, however. Though not sufficient to determine absolute abundances, we claim that the calculations performed here and in [47] are suitable for exploring *changes* in final XRB abundances arising from variations in the input nuclear physics, especially when such nuclear physics variations do not result in changes to the nuclear energy generation rate during the burst (see below).

Table III gives our results from individually varying each reaction with Q < 1 MeV in our network by ± ΔQ, for each of our ten XRB models. Each run involved changing the reaction Q-value of only a single reaction. Since the intent of this work is to identify those reactions whose ΔQ have the largest impact on XRB nucleosynthesis, we have included in Table III only those nuclides that attain a mass fraction $X_f$ > 10$^{-5}$ at the end of the burst, and differ from mass fractions $X_{f,std}$ (calculated with standard rates and Q-values) by at least a factor of two. In addition, as in Table II, we show only isotopes that are stable or with $t_{1/2}$ > 1 hour (all other nuclei are assumed to fully decay at the end of the burst, and are added to the appropriate stable or long-lived daughter nuclei).

From Table III we see that the 'short' model is sensitive to the most input reaction Q-values, with uncertainties ΔQ in eight different reactions affecting the final yield of at least one isotope by at least a factor of two in this model. Final yields in the 'K04', 'hiT', 'lowT' and 'lowZ' models are also sensitive to different input reaction Q-values: uncertainties ΔQ in 5 – 6 reactions affect yields in these models. Conversely, the 'F08', 'hiZ' and 'hiZ2' models are sensitive to the fewest input reaction Q-values in our studies; the ΔQ of only one reaction affects final XRB yields in each of these models. Indeed, the 'F08' model is most robust to the effects of reaction Q-value uncertainties – the only change relative to yields with standard rates (Table II) is in the final yield of $^{45}$Ti, which changes by a factor of two when the Q-value of the $^{45}$Cr(p,γ)$^{46}$Mn reaction is varied. The uncertainty in the Q-value of the $^{64}$Ge(p,γ)$^{65}$As reaction affects by far the most final XRB abundances: isotopic yields from Zn up to as high as Ag are affected in the 'K04', 'lowT', 'long', 'short' and 'lowZ' models.

Although all charged-particle reactions with Q < 1 MeV were varied by ± ΔQ, no (α,γ) or (α,p) reactions appear in Table III. No significant effects on any final XRB yields, in any model, were seen by varying these reactions by their respective ΔQ. This is perhaps expected given that α-induced reactions at (p,γ)-(γ,p) waiting points do not have sufficiently small Q-values. Indeed, Q < 1 MeV for these types of reactions only when another available reaction channel – namely (p,γ) – has a large Q-value. More surprisingly, we find only 15 (p,γ) reactions to have a significant impact on final XRB yields when varied by their respective ΔQ, in any of the ten models. These 15 reactions are summarized in Table IV. The ΔQ of the $^{64}$Ge(p,γ)$^{65}$As reaction affects yields in all but one of our ten models, with a broad range of nuclei often affected, as mentioned above. Uncertainties in the Q-values of the $^{42}$Ti(p,γ)$^{43}$V, $^{46}$Cr(p,γ)$^{47}$Mn, $^{55}$Ni(p,γ)$^{56}$Cu and $^{60}$Zn(p,γ)$^{61}$Ga reactions affect yields in multiple models, but these effects are generally limited to one or two nuclei in the immediate vicinities of these reactions. Uncertainties in the Q-values of over half of the reactions in Table IV affect only a few final yields (1 – 4, by at least a factor of two) in only one of the 10 models examined in this study.

Table IV also indicates those reactions that, in addition to modifying $X_f$ according to the above criteria, modify the nuclear energy generation rate by more than 5% during the burst when varied by ± ΔQ. These effects on the energy generation rate must be interpreted carefully. A one-zone post-processing calculation is not sufficient to predict XRB light curves (since, most notably, a post-processing code cannot self-adjust to allow for variations in the input thermodynamic histories caused by say, changing a reaction rate). However, it should hold that if the Q-value variation of a reaction in a post-processing calculation does *not* affect the nuclear energy generation rate during a burst, then it is unlikely that this ΔQ would strongly affect the XRB light curve predicted by a full hydrodynamic calculation. As can be seen from Table IV, this consideration further emphasizes the crucial Q-value of the $^{64}$Ge(p,γ)$^{65}$As reaction; this Q-value may indeed have strong effects on XRB light curves. However, we stress again that light curves can only be rigorously analyzed in self-consistent hydrodynamic calculations.

IV. CONCLUSIONS

We encourage experimental determinations of the reaction Q-values in Table IV to better constrain the reaction rate equilibria that develop in XRB nucleosynthesis calculations (for $^{30}$S(p,γ)$^{31}$Cl and $^{60}$Zn(p,γ)$^{61}$Ga, the only cases in Table IV for which the masses of all nuclides have been measured, we require the Q-value to a precision better than ±50 keV and ±54 keV, respectively). The question of 'desired precision' is difficult given that most of the Q-values in Table IV are theoretical estimates; however, we find that individually varying the Q-value of each reaction in Table IV by Q ± 0.2×ΔQ keV leads to negligible effects on $X_f$ and nuclear energy generation rate for every model. With regard to the relevant masses, after reviewing measurements of proton-rich isotopes that have been made since the evaluation of Audi et al. [55] (e.g., [58 – 67]) we find that experimental determinations of the masses of $^{26}$P, $^{27}$S, $^{43}$V, $^{46}$Mn, $^{47}$Mn, $^{51}$Co, $^{56}$Cu, $^{65}$As, $^{69}$Br, $^{89}$Ru, $^{90}$Rh, $^{99}$In, $^{106}$Sb and $^{107}$Sb are still lacking. In addition, we require the experimentally-known masses of at least $^{31}$Cl, $^{45}$Cr and $^{61}$Ga to better precision than that given in [55] (±50, 503 and 53 keV, respectively). In particular, we stress the importance of measuring the mass of $^{65}$As ($t_{1/2}$ = 170 ms) (first mentioned in [68, 69] in connection with XRBs) since the uncertainty in the Q-value of the $^{64}$Ge(p,γ)$^{65}$As reaction has by far the largest effects in our XRB models[1].

Reaction Q-values are also vital input for reliable theoretical rate calculations (i.e., 'forward' rate calculations). In Tables 19 – 21 of Parikh et al. [47], reaction rates of importance for XRB studies were identified; mass measurements of the nuclides involved in those reactions with purely theoretical rates are also essential. Of these, we find that experimental determinations of the masses of $^{62}$Ge, $^{65}$As, $^{66}$Se, $^{69}$Br, $^{70}$Kr, $^{84}$Nb, $^{85}$Mo, $^{86}$Tc, $^{87}$Tc, $^{96}$Ag, $^{97}$Cd, $^{103}$Sn and $^{106}$Sb are lacking. Better precision for the experimentally-known masses of $^{71}$Br, $^{83}$Nb and $^{86}$Mo may be required as these are known to only ±568, 315 and 438 keV respectively [55]. We see that the mass of $^{65}$As is critical in this context as well, since variation of the $^{65}$As(p,γ)$^{66}$Se rate led to significant effects in most XRB models of that study. Measurements of the mass of $^{66}$Se ($t_{1/2}$ = 33 ms) as well as the spectroscopy of $^{66}$Se are also urgently needed to improve our knowledge of the important $^{65}$As(p,γ)$^{66}$Se reaction rate in XRBs (see also [48] for further discussion).

---

[1] The mass of $^{64}$Ge has recently been measured to high precision [59, 61]. Our results are unaffected, however, due to the large estimated uncertainty in the mass of $^{65}$As (± 302 keV in [55]).

ACKNOWLEDGEMENTS

We would like to thank the referee for providing constructive comments and helping to improve this manuscript. This work was supported in part by the DFG Cluster of Excellence 'Origin and Structure of the Universe' (www.universecluster.de), the Spanish MEC grant AYA2007-66256, the E.U. FEDER funds, the US DOE under Contract No. DE FG02-97ER41041, and the Swiss NSF grant 200020-122287.

Table I: Summary of the ten XRB scenarios used in our calculations (see text and [47] for more details). Sensitivity to reaction Q-value uncertainties was explored by sampling the parameter space of XRB models in underlying model, peak temperature $T_p$, initial composition $(XYZ)_i$ (where X, Y, Z are $^1$H, $^4$He and metallicity, respectively, by mass), and burst duration $\Delta t$. (Here, we take 'burst duration' as the characteristic timescale of the temperature and density vs. time thermodynamic histories.)

| Model | $T_p$ (GK) | $(XYZ)_i$ | $\Delta t$ (s) | $X_{f,max}$[a] | Endpoint[b] ($X_f > 10^{-2}$) |
|---|---|---|---|---|---|
| K04 | 1.36 | (0.73,0.25,0.02) | ~100 | $^1$H, $^{68}$Ge, $^{72}$Se, $^{64}$Zn, $^{76}$Kr | $^{96}$Ru |
| S01 | 1.91 | (0.718,0.281,0.001) | ~300 | $^{104}$Ag, $^{106}$Cd, $^{105}$Ag, $^{103}$Ag, $^1$H | $^{107}$Cd |
| F08 | 0.99 | (0.40,0.41,0.19) | ~50 | $^{60}$Ni, $^{56}$Ni, $^4$He, $^{28}$Si, $^{12}$C | $^{72}$Se |
| hiT | 2.50 | (0.73,0.25,0.02) | ~100 | $^1$H, $^{72}$Se, $^{68}$Ge, $^{76}$Kr, $^{80}$Sr | $^{103}$Ag |
| lowT | 0.90 | (0.73,0.25,0.02) | ~100 | $^{64}$Zn, $^{68}$Ge, $^1$H, $^{72}$Se, $^{60}$Ni | $^{82}$Sr |
| long | 1.36 | (0.73,0.25,0.02) | ~1000 | $^{68}$Ge, $^{72}$Se, $^{104}$Ag, $^{76}$Kr, $^{103}$Ag | $^{106}$Cd |
| short | 1.36 | (0.73,0.25,0.02) | ~10 | $^1$H, $^{64}$Zn, $^{60}$Ni, $^4$He, $^{68}$Ge | $^{68}$Ge |
| lowZ | 1.36 | (0.7448,0.2551,10$^{-4}$) | ~100 | $^{68}$Ge, $^1$H, $^{72}$Se, $^{64}$Zn, $^{76}$Kr | $^{96}$Ru |
| hiZ | 1.36 | (0.40,0.41,0.19) | ~100 | $^{56}$Ni, $^{60}$Ni, $^{64}$Zn, $^{39}$K, $^{68}$Ge | $^{72}$Se |
| hiZ2 | 1.36 | (0.60,0.21,0.19) | ~100 | $^{60}$Ni, $^{64}$Zn, $^{56}$Ni, $^4$He, $^{68}$Ge | $^{68}$Ge |

Table II: Final abundances (mass fractions) from our one-zone post-processing XRB calculations. See Table I for a summary of the model properties. These final XRB yields were obtained using standard rates in our network (namely, before any Q-values were varied). Only those yields with final mass fractions $X_f > 10^{-10}$ are shown here; all nuclei with $t_{1/2} < 1$ hr have been assumed to fully decay to the nearest stable or long-lived daughter nuclei at the end of the burst.

| Nucleus | Model K04 | S01 | F08 | hiT | lowT | long | short | lowZ | hiZ | hiZ2 |
|---|---|---|---|---|---|---|---|---|---|---|
| $^1$H | $2.0 \times 10^{-1}$ | $7.1 \times 10^{-2}$ | ... | $4.6 \times 10^{-1}$ | $1.5 \times 10^{-1}$ | ... | $4.2 \times 10^{-1}$ | $1.9 \times 10^{-1}$ | ... | ... |
| $^4$He | $2.1 \times 10^{-2}$ | $1.3 \times 10^{-2}$ | $8.5 \times 10^{-2}$ | $1.3 \times 10^{-2}$ | $3.4 \times 10^{-2}$ | $5.1 \times 10^{-3}$ | $6.1 \times 10^{-2}$ | $2.1 \times 10^{-2}$ | $1.8 \times 10^{-2}$ | $1.8 \times 10^{-2}$ |
| $^{12}$C | ... | ... | $4.0 \times 10^{-2}$ | ... | ... | $1.5 \times 10^{-3}$ | ... | ... | $9.5 \times 10^{-3}$ | $9.1 \times 10^{-3}$ |
| $^{13}$C | ... | ... | $5.2 \times 10^{-9}$ | $1.7 \times 10^{-10}$ | ... | ... | $1.2 \times 10^{-10}$ | ... | ... | ... |
| $^{14}$N | $6.6 \times 10^{-7}$ | $6.6 \times 10^{-7}$ | ... | $3.3 \times 10^{-7}$ | $1.2 \times 10^{-6}$ | ... | $5.5 \times 10^{-6}$ | $6.6 \times 10^{-7}$ | ... | ... |
| $^{15}$N | $3.8 \times 10^{-5}$ | $2.9 \times 10^{-5}$ | ... | $1.9 \times 10^{-5}$ | $6.6 \times 10^{-5}$ | ... | $1.8 \times 10^{-4}$ | $3.8 \times 10^{-5}$ | ... | ... |
| $^{16}$O | $3.2 \times 10^{-10}$ | $4.5 \times 10^{-10}$ | $1.8 \times 10^{-4}$ | ... | $2.2 \times 10^{-9}$ | $5.6 \times 10^{-6}$ | $1.7 \times 10^{-9}$ | $3.3 \times 10^{-10}$ | $3.5 \times 10^{-5}$ | $3.3 \times 10^{-5}$ |
| $^{17}$O | $1.6 \times 10^{-9}$ | $1.6 \times 10^{-9}$ | ... | $2.2 \times 10^{-10}$ | $5.8 \times 10^{-9}$ | ... | $1.8 \times 10^{-8}$ | $1.6 \times 10^{-9}$ | ... | ... |
| $^{18}$F | $2.7 \times 10^{-4}$ | $8.2 \times 10^{-5}$ | ... | $1.1 \times 10^{-4}$ | $5.0 \times 10^{-4}$ | ... | $3.9 \times 10^{-3}$ | $2.7 \times 10^{-4}$ | ... | ... |
| $^{19}$F | $1.9 \times 10^{-9}$ | $2.3 \times 10^{-9}$ | ... | $2.6 \times 10^{-10}$ | $6.8 \times 10^{-9}$ | ... | $1.2 \times 10^{-8}$ | $1.9 \times 10^{-9}$ | ... | ... |
| $^{20}$Ne | $4.1 \times 10^{-9}$ | $1.2 \times 10^{-9}$ | $5.0 \times 10^{-4}$ | $1.2 \times 10^{-8}$ | $5.2 \times 10^{-9}$ | $1.6 \times 10^{-5}$ | $2.1 \times 10^{-7}$ | $3.9 \times 10^{-9}$ | $9.3 \times 10^{-5}$ | $9.0 \times 10^{-5}$ |
| $^{21}$Ne | $2.2 \times 10^{-5}$ | $6.9 \times 10^{-6}$ | $7.6 \times 10^{-10}$ | $9.4 \times 10^{-6}$ | $4.2 \times 10^{-5}$ | ... | $2.9 \times 10^{-4}$ | $2.2 \times 10^{-5}$ | ... | ... |
| $^{22}$Na | $3.9 \times 10^{-6}$ | $4.5 \times 10^{-6}$ | ... | $3.3 \times 10^{-7}$ | $2.2 \times 10^{-5}$ | ... | $2.1 \times 10^{-5}$ | $4.1 \times 10^{-6}$ | ... | ... |
| $^{23}$Na | $3.4 \times 10^{-9}$ | $2.3 \times 10^{-9}$ | ... | $4.8 \times 10^{-10}$ | $1.3 \times 10^{-8}$ | ... | $2.7 \times 10^{-8}$ | $3.5 \times 10^{-9}$ | ... | ... |
| $^{24}$Mg | $3.3 \times 10^{-5}$ | $9.1 \times 10^{-6}$ | $4.4 \times 10^{-3}$ | $1.4 \times 10^{-5}$ | $6.3 \times 10^{-5}$ | $1.8 \times 10^{-4}$ | $5.6 \times 10^{-4}$ | $3.3 \times 10^{-5}$ | $7.6 \times 10^{-4}$ | $7.3 \times 10^{-4}$ |
| $^{25}$Mg | $5.4 \times 10^{-5}$ | $1.5 \times 10^{-5}$ | $4.4 \times 10^{-4}$ | $1.9 \times 10^{-5}$ | $1.0 \times 10^{-4}$ | $1.8 \times 10^{-7}$ | $9.2 \times 10^{-4}$ | $5.4 \times 10^{-5}$ | $3.1 \times 10^{-7}$ | $1.8 \times 10^{-6}$ |
| $^{26}$Mg | $5.6 \times 10^{-8}$ | $3.3 \times 10^{-8}$ | $4.4 \times 10^{-5}$ | $2.3 \times 10^{-8}$ | $2.6 \times 10^{-7}$ | $1.0 \times 10^{-8}$ | $8.6 \times 10^{-7}$ | $5.7 \times 10^{-8}$ | $8.1 \times 10^{-7}$ | $7.6 \times 10^{-8}$ |
| $^{26}$Al | ... | ... | $4.5 \times 10^{-3}$ | ... | ... | $3.1 \times 10^{-10}$ | ... | ... | $2.9 \times 10^{-10}$ | $1.1 \times 10^{-8}$ |
| $^{27}$Al | $4.9 \times 10^{-6}$ | $4.4 \times 10^{-6}$ | $1.4 \times 10^{-3}$ | $1.0 \times 10^{-6}$ | $1.7 \times 10^{-5}$ | $1.3 \times 10^{-7}$ | $5.0 \times 10^{-5}$ | $5.0 \times 10^{-6}$ | $1.0 \times 10^{-6}$ | $9.6 \times 10^{-7}$ |
| $^{28}$Si | $3.3 \times 10^{-5}$ | $7.5 \times 10^{-6}$ | $4.1 \times 10^{-2}$ | $1.4 \times 10^{-5}$ | $5.9 \times 10^{-5}$ | $1.9 \times 10^{-3}$ | $6.4 \times 10^{-4}$ | $3.3 \times 10^{-5}$ | $7.2 \times 10^{-3}$ | $7.0 \times 10^{-3}$ |
| $^{29}$Si | $5.5 \times 10^{-5}$ | $1.5 \times 10^{-5}$ | $9.9 \times 10^{-3}$ | $2.3 \times 10^{-5}$ | $1.0 \times 10^{-4}$ | $4.7 \times 10^{-7}$ | $1.1 \times 10^{-3}$ | $5.5 \times 10^{-5}$ | $1.5 \times 10^{-6}$ | $2.4 \times 10^{-5}$ |
| $^{30}$Si | $2.8 \times 10^{-4}$ | $9.1 \times 10^{-5}$ | $2.9 \times 10^{-2}$ | $4.6 \times 10^{-5}$ | $6.3 \times 10^{-4}$ | $2.4 \times 10^{-6}$ | $4.6 \times 10^{-3}$ | $2.8 \times 10^{-4}$ | $3.0 \times 10^{-5}$ | $7.0 \times 10^{-4}$ |
| $^{31}$P | $4.4 \times 10^{-6}$ | $9.4 \times 10^{-7}$ | $7.2 \times 10^{-3}$ | $2.2 \times 10^{-6}$ | $5.2 \times 10^{-6}$ | $5.7 \times 10^{-6}$ | $1.6 \times 10^{-4}$ | $4.4 \times 10^{-6}$ | $9.0 \times 10^{-5}$ | $6.2 \times 10^{-4}$ |
| $^{32}$S | $5.2 \times 10^{-6}$ | $2.7 \times 10^{-7}$ | $9.1 \times 10^{-3}$ | $7.8 \times 10^{-6}$ | $3.1 \times 10^{-6}$ | $9.2 \times 10^{-3}$ | $4.1 \times 10^{-4}$ | $5.0 \times 10^{-6}$ | $3.6 \times 10^{-2}$ | $1.4 \times 10^{-2}$ |
| $^{33}$S | 6.0x10$^{-5}$ | 1.6x10$^{-5}$ | 4.8x10$^{-3}$ | 2.4x10$^{-5}$ | 1.1x10$^{-4}$ | 1.6x10$^{-4}$ | 1.9x10$^{-3}$ | 6.0x10$^{-5}$ | 1.7x10$^{-3}$ | 1.8x10$^{-3}$ |
| $^{34}$S | 2.2x10$^{-4}$ | 7.9x10$^{-5}$ | 2.8x10$^{-2}$ | 5.0x10$^{-5}$ | 5.1x10$^{-4}$ | 3.2x10$^{-5}$ | 5.2x10$^{-3}$ | 2.3x10$^{-4}$ | 6.8x10$^{-3}$ | 2.3x10$^{-3}$ |
| $^{35}$Cl | 1.0x10$^{-7}$ | 8.7x10$^{-8}$ | 9.0x10$^{-3}$ | 1.8x10$^{-8}$ | 3.2x10$^{-7}$ | 2.1x10$^{-4}$ | 2.0x10$^{-6}$ | 1.0x10$^{-7}$ | 1.4x10$^{-2}$ | 1.9x10$^{-3}$ |
| $^{36}$Cl | ... | ... | 6.7x10$^{-9}$ | ... | ... | 1.2x10$^{-7}$ | ... | ... | 4.5x10$^{-5}$ | 1.0x10$^{-6}$ |
| $^{36}$Ar | 1.0x10$^{-5}$ | 6.1x10$^{-7}$ | 7.4x10$^{-3}$ | 9.3x10$^{-6}$ | 8.9x10$^{-6}$ | 4.6x10$^{-3}$ | 6.8x10$^{-4}$ | 1.0x10$^{-5}$ | 1.8x10$^{-2}$ | 1.6x10$^{-3}$ |
| $^{37}$Cl | ... | ... | 3.3x10$^{-9}$ | ... | ... | 7.4x10$^{-9}$ | ... | ... | 1.5x10$^{-7}$ | 3.9x10$^{-9}$ |
| $^{37}$Ar | 7.1x10$^{-5}$ | 2.0x10$^{-5}$ | 5.6x10$^{-4}$ | 2.9x10$^{-5}$ | 1.4x10$^{-4}$ | 4.0x10$^{-5}$ | 2.7x10$^{-3}$ | 7.1x10$^{-5}$ | 5.9x10$^{-3}$ | 2.1x10$^{-4}$ |
| $^{38}$Ar | 1.8x10$^{-4}$ | 4.9x10$^{-5}$ | 2.1x10$^{-2}$ | 7.2x10$^{-5}$ | 3.4x10$^{-4}$ | 2.2x10$^{-5}$ | 8.4x10$^{-3}$ | 1.8x10$^{-4}$ | 2.7x10$^{-2}$ | 1.8x10$^{-3}$ |
| $^{39}$Ar | ... | ... | ... | ... | ... | 2.0x10$^{-10}$ | ... | ... | 6.3x10$^{-7}$ | 6.2x10$^{-10}$ |
| $^{39}$K | 3.5x10$^{-6}$ | 2.9x10$^{-6}$ | 2.6x10$^{-2}$ | 4.6x10$^{-7}$ | 1.3x10$^{-5}$ | 2.0x10$^{-4}$ | 7.7x10$^{-5}$ | 3.6x10$^{-6}$ | 5.7x10$^{-2}$ | 4.1x10$^{-3}$ |
| $^{40}$Ar | ... | ... | ... | ... | ... | ... | ... | ... | 4.0x10$^{-10}$ | ... |
| $^{40}$K | ... | ... | ... | ... | ... | 8.3x10$^{-10}$ | ... | ... | 1.6x10$^{-6}$ | 2.2x10$^{-9}$ |
| $^{40}$Ca | 1.1x10$^{-7}$ | 1.2x10$^{-7}$ | 2.7x10$^{-3}$ | 1.5x10$^{-8}$ | 4.1x10$^{-7}$ | 9.1x10$^{-4}$ | 2.6x10$^{-6}$ | 1.2x10$^{-7}$ | 1.1x10$^{-2}$ | 1.4x10$^{-3}$ |
| $^{41}$K | ... | ... | ... | ... | ... | ... | ... | ... | 7.4x10$^{-9}$ | ... |
| $^{41}$Ca | 3.5x10$^{-5}$ | 9.5x10$^{-6}$ | 1.1x10$^{-6}$ | 1.4x10$^{-5}$ | 6.7x10$^{-5}$ | 3.3x10$^{-7}$ | 1.7x10$^{-3}$ | 3.5x10$^{-5}$ | 4.0x10$^{-4}$ | 2.4x10$^{-6}$ |
| $^{42}$Ca | 9.7x10$^{-6}$ | 1.1x10$^{-5}$ | 5.7x10$^{-3}$ | 5.1x10$^{-7}$ | 5.2x10$^{-5}$ | 5.6x10$^{-7}$ | 1.2x10$^{-4}$ | 1.0x10$^{-5}$ | 6.6x10$^{-3}$ | 3.6x10$^{-4}$ |
| $^{43}$Ca | ... | ... | 5.4x10$^{-6}$ | ... | ... | 2.1x10$^{-8}$ | ... | ... | 3.9x10$^{-5}$ | 4.1x10$^{-6}$ |
| $^{43}$Sc | 6.2x10$^{-8}$ | 5.6x10$^{-8}$ | 4.0x10$^{-3}$ | 2.6x10$^{-8}$ | 2.0x10$^{-7}$ | 4.9x10$^{-7}$ | 2.4x10$^{-6}$ | 6.3x10$^{-8}$ | 8.6x10$^{-3}$ | 9.9x10$^{-4}$ |
| $^{44}$Ca | ... | ... | ... | ... | ... | ... | ... | ... | 1.5x10$^{-9}$ | ... |
| $^{44}$Sc | ... | ... | 1.7x10$^{-8}$ | ... | ... | 2.9x10$^{-10}$ | ... | ... | 3.8x10$^{-7}$ | 2.8x10$^{-8}$ |
| $^{44}$Ti | 2.3x10$^{-5}$ | 3.8x10$^{-6}$ | 6.1x10$^{-4}$ | 1.0x10$^{-5}$ | 3.4x10$^{-5}$ | 3.1x10$^{-5}$ | 1.2x10$^{-3}$ | 2.3x10$^{-5}$ | 1.6x10$^{-3}$ | 8.6x10$^{-5}$ |
| $^{45}$Sc | ... | ... | 3.2x10$^{-7}$ | ... | ... | 1.1x10$^{-9}$ | ... | ... | 2.1x10$^{-7}$ | 5.5x10$^{-10}$ |
| $^{45}$Ti | 3.0x10$^{-7}$ | 3.4x10$^{-7}$ | 1.8x10$^{-4}$ | 3.9x10$^{-8}$ | 1.1x10$^{-6}$ | 1.9x10$^{-8}$ | 7.0x10$^{-6}$ | 3.1x10$^{-7}$ | 2.3x10$^{-6}$ | 6.2x10$^{-8}$ |
| $^{46}$Ti | 9.5x10$^{-5}$ | 3.4x10$^{-5}$ | 1.2x10$^{-2}$ | 1.2x10$^{-5}$ | 2.3x10$^{-4}$ | 3.2x10$^{-9}$ | 2.5x10$^{-3}$ | 9.7x10$^{-5}$ | 1.5x10$^{-2}$ | 5.1x10$^{-4}$ |
| $^{47}$Ti | 1.1x10$^{-5}$ | 2.9x10$^{-6}$ | 1.0x10$^{-3}$ | 4.4x10$^{-6}$ | 2.1x10$^{-5}$ | 6.1x10$^{-7}$ | 5.5x10$^{-4}$ | 1.1x10$^{-5}$ | 7.4x10$^{-3}$ | 1.1x10$^{-4}$ |
| $^{48}$Ti | ... | ... | ... | ... | ... | ... | ... | ... | 1.6x10$^{-10}$ | ... |
| $^{48}$V | ... | ... | 7.6x10$^{-7}$ | ... | ... | 1.8x10$^{-7}$ | ... | ... | 4.5x10$^{-6}$ | 1.9x10$^{-7}$ |
| $^{48}$Cr | 6.6x10$^{-6}$ | 7.5x10$^{-7}$ | 3.0x10$^{-3}$ | 7.1x10$^{-6}$ | 5.3x10$^{-6}$ | 2.0x10$^{-5}$ | 7.4x10$^{-4}$ | 6.4x10$^{-6}$ | 4.7x10$^{-3}$ | 2.5x10$^{-4}$ |
| $^{49}$V | 3.7x10$^{-5}$ | 1.0x10$^{-5}$ | 2.3x10$^{-3}$ | 1.5x10$^{-5}$ | 7.1x10$^{-5}$ | 1.8x10$^{-6}$ | 2.0x10$^{-3}$ | 3.8x10$^{-5}$ | 1.6x10$^{-5}$ | 4.3x10$^{-5}$ |
| $^{50}$V | ... | ... | ... | ... | ... | ... | ... | ... | 1.2x10$^{-10}$ | ... |
| $^{50}$Cr | 8.5x10$^{-5}$ | 2.3x10$^{-5}$ | 9.9x10$^{-3}$ | 3.2x10$^{-5}$ | 1.6x10$^{-4}$ | 4.5x10$^{-7}$ | 4.6x10$^{-3}$ | 8.5x10$^{-5}$ | 1.6x10$^{-2}$ | 5.9x10$^{-4}$ |
| $^{51}$V | ... | ... | 4.4x10$^{-10}$ | ... | ... | 8.6x10$^{-10}$ | ... | ... | 8.3x10$^{-9}$ | 4.1x10$^{-10}$ |
| $^{51}$Cr | 6.5x10$^{-7}$ | 6.8x10$^{-7}$ | 1.6x10$^{-2}$ | 1.7x10$^{-7}$ | 2.2x10$^{-6}$ | 2.8x10$^{-5}$ | 2.5x10$^{-5}$ | 6.7x10$^{-7}$ | 2.7x10$^{-2}$ | 1.6x10$^{-3}$ |
| $^{52}$Cr | ... | ... | 1.8x10$^{-10}$ | ... | ... | 5.7x10$^{-9}$ | ... | ... | 1.7x10$^{-9}$ | 2.0x10$^{-10}$ |
| $^{52}$Mn | ... | ... | 4.2x10$^{-6}$ | ... | ... | 3.6x10$^{-6}$ | ... | ... | 1.0x10$^{-5}$ | 1.5x10$^{-6}$ |
| $^{52}$Fe | 8.7x10$^{-6}$ | 1.3x10$^{-5}$ | 6.4x10$^{-3}$ | 1.5x10$^{-6}$ | 3.3x10$^{-5}$ | 1.5x10$^{-4}$ | 2.5x10$^{-4}$ | 9.0x10$^{-6}$ | 4.5x10$^{-3}$ | 7.7x10$^{-4}$ |
| $^{53}$Mn | 2.8x10$^{-5}$ | 9.3x10$^{-6}$ | 1.1x10$^{-3}$ | 1.1x10$^{-5}$ | 5.6x10$^{-5}$ | 9.4x10$^{-7}$ | 1.5x10$^{-3}$ | 2.8x10$^{-5}$ | 5.3x10$^{-6}$ | 1.9x10$^{-5}$ |
| $^{54}$Fe | 6.3x10$^{-5}$ | 1.8x10$^{-5}$ | 4.2x10$^{-3}$ | 2.5x10$^{-5}$ | 1.2x10$^{-4}$ | 1.3x10$^{-6}$ | 3.6x10$^{-3}$ | 6.3x10$^{-5}$ | 8.2x10$^{-3}$ | 3.4x10$^{-4}$ |
| $^{55}$Fe | ... | ... | 4.7x10$^{-6}$ | ... | ... | 4.2x10$^{-7}$ | ... | ... | 3.1x10$^{-5}$ | 1.2x10$^{-6}$ |
| $^{55}$Co | 2.2x10$^{-5}$ | 1.7x10$^{-5}$ | 1.5x10$^{-2}$ | 1.5x10$^{-6}$ | 9.6x10$^{-5}$ | 3.9x10$^{-5}$ | 4.0x10$^{-4}$ | 2.3x10$^{-5}$ | 3.1x10$^{-2}$ | 1.3x10$^{-3}$ |
| $^{56}$Fe | ... | ... | ... | ... | ... | 1.4x10$^{-10}$ | ... | ... | 1.6x10$^{-10}$ | ... |
| $^{56}$Co | ... | ... | 4.6x10$^{-6}$ | ... | ... | 2.7x10$^{-6}$ | ... | ... | 3.3x10$^{-5}$ | 5.6x10$^{-6}$ |
| $^{56}$Ni | 3.2x10$^{-5}$ | 3.1x10$^{-5}$ | 1.3x10$^{-1}$ | 4.1x10$^{-6}$ | 1.3x10$^{-4}$ | 2.1x10$^{-3}$ | 8.8x10$^{-4}$ | 3.3x10$^{-5}$ | 2.7x10$^{-1}$ | 5.1x10$^{-2}$ |
| $^{57}$Co | ... | ... | 5.7x10$^{-9}$ | ... | ... | 1.6x10$^{-9}$ | ... | ... | 5.0x10$^{-9}$ | 4.3x10$^{-10}$ |
| $^{57}$Ni | 3.2x10$^{-5}$ | 1.8x10$^{-5}$ | 3.8x10$^{-5}$ | 1.1x10$^{-5}$ | 7.2x10$^{-5}$ | 3.0x10$^{-7}$ | 1.7x10$^{-3}$ | 3.2x10$^{-5}$ | 8.0x10$^{-6}$ | 9.5x10$^{-7}$ |
| $^{58}$Ni | 5.5x10$^{-5}$ | 1.9x10$^{-5}$ | 3.4x10$^{-4}$ | 2.2x10$^{-5}$ | 1.1x10$^{-4}$ | 3.2x10$^{-8}$ | 3.3x10$^{-3}$ | 5.5x10$^{-5}$ | 8.7x10$^{-5}$ | 5.1x10$^{-6}$ |
| $^{59}$Ni | 1.3x10$^{-4}$ | 4.6x10$^{-5}$ | 3.8x10$^{-3}$ | 4.9x10$^{-5}$ | 2.6x10$^{-4}$ | 2.9x10$^{-7}$ | 8.2x10$^{-3}$ | 1.3x10$^{-4}$ | 3.8x10$^{-3}$ | 9.9x10$^{-5}$ |
| $^{60}$Ni | 7.0x10$^{-3}$ | 7.7x10$^{-3}$ | 3.8x10$^{-1}$ | 4.6x10$^{-4}$ | 5.2x10$^{-2}$ | 8.5x10$^{-3}$ | 1.0x10$^{-1}$ | 7.3x10$^{-3}$ | 2.6x10$^{-1}$ | 7.0x10$^{-1}$ |
| $^{61}$Ni | ... | ... | 3.4x10$^{-8}$ | ... | ... | 7.1x10$^{-6}$ | ... | ... | 6.8x10$^{-8}$ | 4.5x10$^{-7}$ |
| $^{61}$Cu | 2.3x10$^{-5}$ | 2.7x10$^{-5}$ | 2.1x10$^{-4}$ | 4.0x10$^{-6}$ | 1.1x10$^{-4}$ | 1.5x10$^{-4}$ | 6.5x10$^{-4}$ | 2.3x10$^{-5}$ | 4.6x10$^{-5}$ | 3.6x10$^{-4}$ |
| $^{62}$Ni | ... | ... | 5.7x10$^{-8}$ | ... | ... | 1.7x10$^{-7}$ | ... | ... | 9.9x10$^{-8}$ | 2.1x10$^{-7}$ |
| $^{62}$Zn | 6.2x10$^{-5}$ | 2.3x10$^{-5}$ | 9.7x10$^{-5}$ | 3.2x10$^{-5}$ | 1.6x10$^{-4}$ | 8.3x10$^{-6}$ | 3.8x10$^{-3}$ | 6.1x10$^{-5}$ | 5.2x10$^{-5}$ | 1.2x10$^{-4}$ |
| $^{63}$Cu | 9.6x10$^{-5}$ | 5.6x10$^{-5}$ | 5.1x10$^{-4}$ | 3.9x10$^{-5}$ | 3.1x10$^{-4}$ | 1.2x10$^{-5}$ | 5.3x10$^{-3}$ | 9.5x10$^{-5}$ | 3.0x10$^{-4}$ | 8.7x10$^{-4}$ |
| $^{64}$Zn | 7.1x10$^{-2}$ | 1.0x10$^{-2}$ | 3.4x10$^{-2}$ | 9.1x10$^{-3}$ | 3.8x10$^{-1}$ | 1.8x10$^{-2}$ | 3.1x10$^{-1}$ | 7.5x10$^{-2}$ | 6.2x10$^{-2}$ | 1.6x10$^{-1}$ |
| $^{65}$Cu | ... | ... | ... | ... | ... | 2.5x10$^{-9}$ | ... | ... | ... | ... |
| $^{65}$Zn | 6.7x10$^{-5}$ | 5.4x10$^{-5}$ | 5.2x10$^{-5}$ | 3.9x10$^{-6}$ | 6.6x10$^{-4}$ | 2.8x10$^{-4}$ | 1.6x10$^{-4}$ | 7.2x10$^{-5}$ | 1.9x10$^{-5}$ | 1.9x10$^{-4}$ |
| $^{66}$Zn | ... | ... | ... | ... | ... | 5.5x10$^{-8}$ | ... | ... | 7.9x10$^{-10}$ | 7.6x10$^{-10}$ |
| $^{66}$Ga | ... | 6.7x10$^{-10}$ | 9.3x10$^{-8}$ | ... | 1.9x10$^{-9}$ | 5.3x10$^{-6}$ | ... | ... | 8.3x10$^{-7}$ | 8.8x10$^{-7}$ |
| $^{66}$Ge | 1.1x10$^{-4}$ | 1.0x10$^{-4}$ | 3.9x10$^{-5}$ | 5.9x10$^{-6}$ | 1.1x10$^{-3}$ | 6.1x10$^{-5}$ | 2.4x10$^{-4}$ | 1.2x10$^{-4}$ | 1.1x10$^{-4}$ | 1.3x10$^{-4}$ |
| $^{67}$Zn | ... | ... | ... | ... | ... | 1.1x10$^{-8}$ | ... | ... | ... | 6.3x10$^{-10}$ |
| $^{67}$Ga | 2.1x10$^{-4}$ | 1.0x10$^{-4}$ | 7.9x10$^{-5}$ | 2.5x10$^{-5}$ | 1.6x10$^{-3}$ | 1.9x10$^{-5}$ | 8.1x10$^{-4}$ | 2.3x10$^{-4}$ | 3.2x10$^{-5}$ | 3.6x10$^{-4}$ |
| $^{68}$Zn | ... | ... | ... | ... | ... | 2.0x10$^{-7}$ | ... | ... | ... | ... |
| $^{68}$Ga | ... | ... | 1.8x10$^{-10}$ | ... | ... | 3.0x10$^{-6}$ | ... | ... | 8.2x10$^{-9}$ | 2.5x10$^{-9}$ |
| $^{68}$Ge | 2.0x10$^{-1}$ | 1.0x10$^{-2}$ | 2.3x10$^{-2}$ | 5.8x10$^{-2}$ | 1.9x10$^{-1}$ | 1.6x10$^{-1}$ | 2.3x10$^{-2}$ | 2.1x10$^{-1}$ | 4.1x10$^{-2}$ | 1.6x10$^{-2}$ |
| $^{69}$Ga | ... | ... | ... | ... | ... | 7.6x10$^{-6}$ | ... | ... | 7.0x10$^{-10}$ | 6.4x10$^{-10}$ |
| $^{69}$Ge | 1.1x10$^{-3}$ | 3.0x10$^{-4}$ | 1.3x10$^{-4}$ | 1.1x10$^{-4}$ | 2.0x10$^{-3}$ | 5.8x10$^{-3}$ | 5.4x10$^{-5}$ | 1.1x10$^{-3}$ | 9.1x10$^{-5}$ | 1.1x10$^{-4}$ |
| $^{70}$Ge | 6.7x10$^{-4}$ | 1.6x10$^{-4}$ | 3.6x10$^{-5}$ | 7.7x10$^{-5}$ | 1.1x10$^{-3}$ | 4.6x10$^{-4}$ | 3.6x10$^{-5}$ | 6.9x10$^{-4}$ | 8.8x10$^{-5}$ | 1.2x10$^{-5}$ |
| $^{71}$Ga | ... | ... | ... | ... | ... | 4.9x10$^{-10}$ | ... | ... | ... | ... |
| $^{71}$Ge | ... | ... | ... | ... | ... | 1.8x10$^{-6}$ | ... | ... | 7.6x10$^{-10}$ | 5.2x10$^{-10}$ |
| $^{71}$As | 1.5x10$^{-3}$ | 3.7x10$^{-4}$ | 9.4x10$^{-5}$ | 2.3x10$^{-4}$ | 2.5x10$^{-3}$ | 1.0x10$^{-3}$ | 8.8x10$^{-5}$ | 1.6x10$^{-3}$ | 2.8x10$^{-5}$ | 4.1x10$^{-5}$ |
| $^{72}$Ge | ... | ... | ... | ... | ... | 3.1x10$^{-7}$ | ... | ... | ... | ... |
| $^{72}$As | ... | ... | 1.3x10$^{-8}$ | ... | ... | 9.6x10$^{-5}$ | ... | ... | 2.7x10$^{-7}$ | 1.9x10$^{-8}$ |
| $^{72}$Se | 1.3x10$^{-1}$ | 7.2x10$^{-3}$ | 1.4x10$^{-2}$ | 7.0x10$^{-2}$ | 7.4x10$^{-2}$ | 1.2x10$^{-1}$ | 1.1x10$^{-3}$ | 1.3x10$^{-1}$ | 1.6x10$^{-2}$ | 1.4x10$^{-3}$ |
| $^{73}$Ge | ... | ... | ... | ... | ... | 4.0x10$^{-9}$ | ... | ... | ... | ... |
| $^{73}$As | ... | ... | 1.2x10$^{-9}$ | ... | ... | 1.1x10$^{-4}$ | ... | ... | 8.5x10$^{-9}$ | 1.8x10$^{-9}$ |
| $^{73}$Se | 2.1x10$^{-3}$ | 6.8x10$^{-4}$ | 1.4x10$^{-4}$ | 3.8x10$^{-4}$ | 2.3x10$^{-3}$ | 5.0x10$^{-3}$ | 6.9x10$^{-6}$ | 2.2x10$^{-3}$ | 4.2x10$^{-5}$ | 1.1x10$^{-5}$ |
| $^{74}$Se | 1.7x10$^{-3}$ | 5.4x10$^{-4}$ | 6.5x10$^{-5}$ | 3.3x10$^{-4}$ | 1.8x10$^{-3}$ | 1.1x10$^{-3}$ | 5.2x10$^{-6}$ | 1.7x10$^{-3}$ | 7.1x10$^{-5}$ | 4.1x10$^{-6}$ |
| $^{75}$As | ... | ... | ... | ... | ... | 7.0x10$^{-10}$ | ... | ... | ... | ... |
| $^{75}$Se | ... | ... | 3.9x10$^{-9}$ | ... | ... | 2.7x10$^{-5}$ | ... | ... | 1.3x10$^{-8}$ | 5.1x10$^{-9}$ |
| $^{75}$Br | 3.2x10$^{-3}$ | 9.0x10$^{-4}$ | 1.2x10$^{-4}$ | 7.6x10$^{-4}$ | 3.2x10$^{-3}$ | 3.6x10$^{-4}$ | 9.5x10$^{-6}$ | 3.2x10$^{-3}$ | 1.6x10$^{-5}$ | 9.1x10$^{-6}$ |
| $^{76}$Se | ... | ... | ... | ... | ... | 5.2x10$^{-6}$ | ... | ... | 5.2x10$^{-10}$ | ... |
| $^{76}$Br | ... | 2.8x10$^{-10}$ | 1.7x10$^{-7}$ | ... | ... | 9.3x10$^{-4}$ | ... | ... | 1.4x10$^{-6}$ | 6.2x10$^{-8}$ |
| $^{76}$Kr | 7.4x10$^{-2}$ | 4.9x10$^{-3}$ | 4.3x10$^{-3}$ | 4.8x10$^{-2}$ | 3.1x10$^{-2}$ | 7.7x10$^{-2}$ | 5.5x10$^{-5}$ | 7.4x10$^{-2}$ | 2.8x10$^{-3}$ | 1.5x10$^{-4}$ |
| $^{77}$Se | ... | ... | ... | ... | ... | 5.7x10$^{-7}$ | ... | ... | ... | ... |
| $^{77}$Br | ... | 1.2x10$^{-10}$ | 7.8x10$^{-9}$ | ... | ... | 4.3x10$^{-4}$ | ... | ... | 3.1x10$^{-8}$ | 2.9x10$^{-9}$ |
| $^{77}$Kr | 4.0x10$^{-3}$ | 1.3x10$^{-3}$ | 9.4x10$^{-5}$ | 9.7x10$^{-4}$ | 3.0x10$^{-3}$ | 4.1x10$^{-3}$ | 1.0x10$^{-6}$ | 4.0x10$^{-3}$ | 2.2x10$^{-5}$ | 2.6x10$^{-6}$ |

| Nuclide | | | | | | | | | | |
|---|---|---|---|---|---|---|---|---|---|---|
| $^{78}$Kr | 4.8x10$^{-3}$ | 1.7x10$^{-3}$ | 6.3x10$^{-5}$ | 1.3x10$^{-3}$ | 3.6x10$^{-3}$ | 8.1x10$^{-4}$ | 9.6x10$^{-7}$ | 4.9x10$^{-3}$ | 2.0x10$^{-6}$ | 1.2x10$^{-6}$ |
| $^{79}$Br | ... | ... | ... | ... | ... | 3.4x10$^{-7}$ | ... | ... | ... | ... |
| $^{79}$Kr | 4.8x10$^{-3}$ | 1.6x10$^{-3}$ | 4.6x10$^{-5}$ | 1.3x10$^{-3}$ | 3.4x10$^{-3}$ | 4.5x10$^{-4}$ | 7.3x10$^{-7}$ | 4.8x10$^{-3}$ | 6.9x10$^{-6}$ | 8.9x10$^{-7}$ |
| $^{80}$Kr | 2.0x10$^{-8}$ | 2.0x10$^{-7}$ | 6.5x10$^{-7}$ | 9.0x10$^{-10}$ | 4.0x10$^{-8}$ | 4.2x10$^{-3}$ | ... | 2.1x10$^{-8}$ | 2.6x10$^{-6}$ | 9.0x10$^{-8}$ |
| $^{80}$Sr | 4.2x10$^{-2}$ | 4.9x10$^{-3}$ | 1.2x10$^{-3}$ | 3.1x10$^{-2}$ | 1.5x10$^{-2}$ | 4.0x10$^{-2}$ | 2.9x10$^{-6}$ | 4.1x10$^{-2}$ | 5.0x10$^{-4}$ | 2.0x10$^{-5}$ |
| $^{81}$Kr | ... | ... | ... | ... | ... | 4.6x10$^{-5}$ | ... | ... | 3.7x10$^{-10}$ | ... |
| $^{81}$Rb | 9.0x10$^{-3}$ | 2.7x10$^{-3}$ | 6.2x10$^{-5}$ | 3.0x10$^{-3}$ | 4.8x10$^{-3}$ | 6.3x10$^{-3}$ | 1.4x10$^{-7}$ | 9.1x10$^{-3}$ | 2.1x10$^{-5}$ | 1.0x10$^{-6}$ |
| $^{82}$Kr | ... | ... | ... | ... | ... | 8.9x10$^{-9}$ | ... | ... | ... | ... |
| $^{82}$Rb | ... | 4.4x10$^{-10}$ | ... | ... | ... | 6.2x10$^{-7}$ | ... | ... | ... | ... |
| $^{82}$Sr | 2.3x10$^{-2}$ | 6.9x10$^{-3}$ | 8.9x10$^{-5}$ | 9.9x10$^{-3}$ | 1.1x10$^{-2}$ | 2.2x10$^{-3}$ | 1.5x10$^{-7}$ | 2.3x10$^{-2}$ | 1.7x10$^{-6}$ | 7.4x10$^{-7}$ |
| $^{83}$Kr | ... | ... | ... | ... | ... | 3.0x10$^{-10}$ | ... | ... | ... | ... |
| $^{83}$Rb | ... | 3.0x10$^{-10}$ | ... | ... | ... | 8.4x10$^{-6}$ | ... | ... | ... | ... |
| $^{83}$Sr | 1.5x10$^{-2}$ | 4.6x10$^{-3}$ | 5.6x10$^{-5}$ | 7.6x10$^{-3}$ | 6.4x10$^{-3}$ | 2.1x10$^{-3}$ | 5.0x10$^{-8}$ | 1.5x10$^{-2}$ | 1.6x10$^{-6}$ | 5.7x10$^{-7}$ |
| $^{84}$Sr | 1.5x10$^{-2}$ | 4.3x10$^{-3}$ | 5.2x10$^{-5}$ | 9.1x10$^{-3}$ | 5.4x10$^{-3}$ | 4.4x10$^{-3}$ | 2.6x10$^{-8}$ | 1.5x10$^{-2}$ | 9.7x10$^{-6}$ | 3.1x10$^{-7}$ |
| $^{85}$Rb | ... | ... | ... | ... | ... | 9.5x10$^{-9}$ | ... | ... | ... | ... |
| $^{85}$Sr | 1.5x10$^{-10}$ | 7.0x10$^{-9}$ | 5.8x10$^{-10}$ | ... | 3.8x10$^{-10}$ | 2.2x10$^{-4}$ | ... | 1.6x10$^{-10}$ | 8.3x10$^{-9}$ | 2.5x10$^{-10}$ |
| $^{85}$Y | 1.6x10$^{-2}$ | 3.8x10$^{-3}$ | 7.6x10$^{-5}$ | 1.1x10$^{-2}$ | 4.3x10$^{-3}$ | 6.8x10$^{-3}$ | 1.5x10$^{-8}$ | 1.5x10$^{-2}$ | 3.9x10$^{-5}$ | 1.5x10$^{-6}$ |
| $^{86}$Sr | ... | ... | ... | ... | ... | 2.7x10$^{-7}$ | ... | ... | ... | ... |
| $^{86}$Y | 1.4x10$^{-9}$ | 2.6x10$^{-8}$ | 9.2x10$^{-10}$ | ... | 2.0x10$^{-9}$ | 4.8x10$^{-5}$ | ... | 1.4x10$^{-9}$ | 8.8x10$^{-10}$ | ... |
| $^{86}$Zr | 1.9x10$^{-2}$ | 4.4x10$^{-3}$ | 8.9x10$^{-5}$ | 1.4x10$^{-2}$ | 4.0x10$^{-3}$ | 4.9x10$^{-3}$ | 7.2x10$^{-9}$ | 1.8x10$^{-2}$ | 4.6x10$^{-6}$ | 5.9x10$^{-7}$ |
| $^{87}$Sr | ... | ... | ... | ... | ... | 1.9x10$^{-7}$ | ... | ... | ... | ... |
| $^{87}$Y | 2.8x10$^{-9}$ | 5.2x10$^{-8}$ | 3.4x10$^{-9}$ | 1.6x10$^{-10}$ | 2.8x10$^{-9}$ | 2.1x10$^{-4}$ | ... | 2.8x10$^{-9}$ | 8.9x10$^{-9}$ | 2.9x10$^{-10}$ |
| $^{87}$Zr | 1.5x10$^{-2}$ | 3.5x10$^{-3}$ | 8.2x10$^{-5}$ | 1.2x10$^{-2}$ | 2.4x10$^{-3}$ | 2.5x10$^{-3}$ | 3.1x10$^{-9}$ | 1.4x10$^{-2}$ | 9.8x10$^{-6}$ | 4.0x10$^{-7}$ |
| $^{88}$Y | ... | ... | ... | ... | ... | 1.1x10$^{-7}$ | ... | ... | ... | ... |
| $^{88}$Zr | 1.0x10$^{-2}$ | 2.8x10$^{-3}$ | 8.3x10$^{-5}$ | 8.2x10$^{-3}$ | 1.5x10$^{-3}$ | 1.1x10$^{-2}$ | 1.2x10$^{-9}$ | 9.8x10$^{-3}$ | 3.9x10$^{-5}$ | 5.5x10$^{-7}$ |
| $^{89}$Y | ... | ... | ... | ... | ... | 9.4x10$^{-7}$ | ... | ... | ... | ... |
| $^{89}$Zr | 2.1x10$^{-8}$ | 2.1x10$^{-7}$ | 8.4x10$^{-9}$ | 2.0x10$^{-9}$ | 9.4x10$^{-9}$ | 9.1x10$^{-4}$ | ... | 2.1x10$^{-8}$ | 3.8x10$^{-8}$ | 9.2x10$^{-10}$ |
| $^{89}$Nb | 1.1x10$^{-2}$ | 2.4x10$^{-3}$ | 6.8x10$^{-5}$ | 1.1x10$^{-2}$ | 1.0x10$^{-3}$ | 6.2x10$^{-3}$ | 4.5x10$^{-10}$ | 1.1x10$^{-2}$ | 1.7x10$^{-5}$ | 5.0x10$^{-7}$ |
| $^{90}$Zr | ... | 1.2x10$^{-10}$ | ... | ... | ... | 8.8x10$^{-6}$ | ... | ... | ... | ... |
| $^{90}$Nb | 2.8x10$^{-7}$ | 1.0x10$^{-6}$ | 1.4x10$^{-8}$ | 5.0x10$^{-8}$ | 5.9x10$^{-8}$ | 1.4x10$^{-3}$ | ... | 2.8x10$^{-7}$ | 2.9x10$^{-8}$ | 6.7x10$^{-10}$ |
| $^{90}$Mo | 1.3x10$^{-2}$ | 2.6x10$^{-3}$ | 4.7x10$^{-5}$ | 1.4x10$^{-2}$ | 7.0x10$^{-4}$ | 4.2x10$^{-2}$ | 1.3x10$^{-10}$ | 1.3x10$^{-2}$ | 1.3x10$^{-5}$ | 3.4x10$^{-7}$ |
| $^{91}$Nb | 1.1x10$^{-2}$ | 2.7x10$^{-3}$ | 2.3x10$^{-5}$ | 1.2x10$^{-2}$ | 4.1x10$^{-4}$ | 6.2x10$^{-3}$ | ... | 1.1x10$^{-2}$ | 3.6x10$^{-6}$ | 2.2x10$^{-8}$ |
| $^{92}$Mo | 1.0x10$^{-2}$ | 2.7x10$^{-3}$ | 7.1x10$^{-6}$ | 1.1x10$^{-2}$ | 2.3x10$^{-4}$ | 2.4x10$^{-3}$ | ... | 1.0x10$^{-2}$ | 9.9x10$^{-7}$ | 7.8x10$^{-9}$ |
| $^{93}$Mo | 9.6x10$^{-8}$ | 5.5x10$^{-7}$ | 3.5x10$^{-10}$ | 2.5x10$^{-8}$ | 3.0x10$^{-9}$ | 3.0x10$^{-4}$ | ... | 9.4x10$^{-8}$ | 3.7x10$^{-9}$ | ... |
| $^{93}$Tc | 9.7x10$^{-3}$ | 2.4x10$^{-3}$ | 3.2x10$^{-6}$ | 1.3x10$^{-2}$ | 9.1x10$^{-5}$ | 4.7x10$^{-3}$ | ... | 9.6x10$^{-3}$ | 2.1x10$^{-6}$ | 1.4x10$^{-8}$ |
| $^{94}$Mo | 1.0x10$^{-2}$ | 3.3x10$^{-3}$ | 6.0x10$^{-7}$ | 1.5x10$^{-2}$ | 3.2x10$^{-5}$ | 5.5x10$^{-2}$ | ... | 1.0x10$^{-2}$ | 2.3x10$^{-6}$ | 8.0x10$^{-9}$ |
| $^{95}$Mo | ... | ... | ... | ... | ... | 2.2x10$^{-6}$ | ... | ... | ... | ... |
| $^{95}$Tc | 6.1x10$^{-8}$ | 4.3x10$^{-7}$ | ... | 2.5x10$^{-8}$ | 1.7x10$^{-10}$ | 6.0x10$^{-4}$ | ... | 6.0x10$^{-8}$ | 1.0x10$^{-10}$ | ... |
| $^{95}$Ru | 8.9x10$^{-3}$ | 3.7x10$^{-3}$ | 6.6x10$^{-8}$ | 1.5x10$^{-2}$ | 9.7x10$^{-6}$ | 8.7x10$^{-3}$ | ... | 8.9x10$^{-3}$ | 1.4x10$^{-7}$ | 4.9x10$^{-10}$ |
| $^{96}$Ru | 1.2x10$^{-2}$ | 5.0x10$^{-3}$ | 6.1x10$^{-9}$ | 2.6x10$^{-2}$ | 2.1x10$^{-6}$ | 3.8x10$^{-3}$ | ... | 1.2x10$^{-2}$ | 7.7x10$^{-9}$ | 1.7x10$^{-10}$ |
| $^{97}$Tc | ... | ... | ... | ... | ... | 6.6x10$^{-6}$ | ... | ... | ... | ... |
| $^{97}$Ru | 3.7x10$^{-3}$ | 2.6x10$^{-3}$ | 6.1x10$^{-10}$ | 9.8x10$^{-3}$ | 8.1x10$^{-8}$ | 1.7x10$^{-2}$ | ... | 3.7x10$^{-3}$ | 2.1x10$^{-7}$ | 1.7x10$^{-10}$ |
| $^{98}$Ru | 4.5x10$^{-3}$ | 3.3x10$^{-3}$ | ... | 1.4x10$^{-2}$ | 1.1x10$^{-8}$ | 4.0x10$^{-2}$ | ... | 4.3x10$^{-3}$ | 6.2x10$^{-8}$ | ... |
| $^{99}$Ru | ... | ... | ... | ... | ... | 5.8x10$^{-7}$ | ... | ... | ... | ... |
| $^{99}$Rh | 3.5x10$^{-3}$ | 5.7x10$^{-3}$ | ... | 1.6x10$^{-2}$ | ... | 7.7x10$^{-3}$ | ... | 3.4x10$^{-3}$ | 1.9x10$^{-9}$ | ... |
| $^{100}$Ru | ... | ... | ... | ... | ... | 4.4x10$^{-8}$ | ... | ... | ... | ... |
| $^{100}$Rh | 6.4x10$^{-10}$ | 1.1x10$^{-8}$ | ... | 1.2x10$^{-9}$ | ... | 1.2x10$^{-5}$ | ... | 6.0x10$^{-10}$ | ... | ... |
| $^{100}$Pd | 2.5x10$^{-3}$ | 4.2x10$^{-3}$ | ... | 1.4x10$^{-2}$ | ... | 7.6x10$^{-3}$ | ... | 2.4x10$^{-3}$ | 2.0x10$^{-9}$ | ... |
| $^{101}$Ru | ... | ... | ... | ... | ... | 3.8x10$^{-10}$ | ... | ... | ... | ... |
| $^{101}$Rh | 3.8x10$^{-10}$ | 2.2x10$^{-8}$ | ... | 1.0x10$^{-9}$ | ... | 1.8x10$^{-4}$ | ... | ... | ... | ... |
| $^{101}$Pd | 1.9x10$^{-3}$ | 8.4x10$^{-3}$ | ... | 1.6x10$^{-2}$ | ... | 2.7x10$^{-2}$ | ... | 1.8x10$^{-3}$ | 4.2x10$^{-9}$ | ... |
| $^{102}$Pd | 1.3x10$^{-3}$ | 1.7x10$^{-2}$ | ... | 1.8x10$^{-2}$ | ... | 3.4x10$^{-2}$ | ... | 1.2x10$^{-3}$ | 6.0x10$^{-10}$ | ... |
| $^{103}$Rh | ... | ... | ... | ... | ... | 7.2x10$^{-7}$ | ... | ... | ... | ... |
| $^{103}$Pd | 5.4x10$^{-9}$ | 5.3x10$^{-6}$ | ... | 5.1x10$^{-8}$ | ... | 4.4x10$^{-3}$ | ... | 4.9x10$^{-9}$ | ... | ... |
| $^{103}$Ag | 7.6x10$^{-4}$ | 7.8x10$^{-2}$ | ... | 1.7x10$^{-2}$ | ... | 6.0x10$^{-2}$ | ... | 7.1x10$^{-4}$ | ... | ... |
| $^{104}$Pd | ... | 9.3x10$^{-7}$ | ... | 7.4x10$^{-10}$ | ... | 9.9x10$^{-4}$ | ... | ... | ... | ... |
| $^{104}$Ag | 2.2x10$^{-4}$ | 3.3x10$^{-1}$ | ... | 8.4x10$^{-3}$ | ... | 9.7x10$^{-2}$ | ... | 2.0x10$^{-4}$ | ... | ... |
| $^{105}$Pd | ... | ... | ... | ... | ... | 3.3x10$^{-7}$ | ... | ... | ... | ... |
| $^{105}$Ag | 1.4x10$^{-5}$ | 8.5x10$^{-2}$ | ... | 7.3x10$^{-4}$ | ... | 4.2x10$^{-2}$ | ... | 1.3x10$^{-5}$ | ... | ... |
| $^{106}$Cd | 3.8x10$^{-6}$ | 2.4x10$^{-1}$ | ... | 3.5x10$^{-4}$ | ... | 1.3x10$^{-2}$ | ... | 3.4x10$^{-6}$ | ... | ... |
| $^{107}$Ag | ... | 4.5x10$^{-8}$ | ... | ... | ... | 2.4x10$^{-6}$ | ... | ... | ... | ... |
| $^{107}$Cd | 3.9x10$^{-8}$ | 2.8x10$^{-2}$ | ... | 5.2x10$^{-6}$ | ... | 8.2x10$^{-4}$ | ... | 3.5x10$^{-8}$ | ... | ... |
| $^{108}$Cd | 2.0x10$^{-9}$ | 1.7x10$^{-3}$ | ... | 5.5x10$^{-7}$ | ... | 1.5x10$^{-5}$ | ... | 1.8x10$^{-9}$ | ... | ... |
| $^{109}$Cd | ... | 7.8x10$^{-9}$ | ... | ... | ... | 4.3x10$^{-8}$ | ... | ... | ... | ... |
| $^{109}$In | ... | 2.9x10$^{-4}$ | ... | 5.3x10$^{-8}$ | ... | 3.8x10$^{-6}$ | ... | ... | ... | ... |
| $^{110}$Sn | ... | 3.5x10$^{-4}$ | ... | 6.5x10$^{-9}$ | ... | ... | ... | ... | ... | ... |
| $^{111}$In | ... | 3.9x10$^{-5}$ | ... | 5.1x10$^{-10}$ | ... | ... | ... | ... | ... | ... |
| $^{112}$Sn | ... | 2.7x10$^{-6}$ | ... | ... | ... | ... | ... | ... | ... | ... |
| $^{113}$Sn | ... | 2.8x10$^{-7}$ | ... | ... | ... | ... | ... | ... | ... | ... |

Table III: Final abundance ratios $X_f/X_{f,std}$ resulting from reaction Q-value variations by $Q \pm \Delta Q$ (see Table IV or [55] for values), for each of the 10 models. Only the most significant abundance changes ($X_f/X_{f,std} > 2$ or $< 0.5$, for $X_f > 10^{-5}$) are listed here.

| Model | Reaction | isotope | Q + ΔQ | Q - ΔQ |
|---|---|---|---|---|
| K04 | $^{26}P(p,\gamma)^{27}S$ | $^{25}Mg$ | 0.35 | … |
| | $^{46}Cr(p,\gamma)^{47}Mn$ | $^{46}Ti$ | 0.23 | … |
| | $^{55}Ni(p,\gamma)^{56}Cu$ | $^{55}Co$ | … | 4.0 |
| | $^{60}Zn(p,\gamma)^{61}Ga$ | $^{60}Ni$ | 0.47 | … |
| | $^{64}Ge(p,\gamma)^{65}As$ | $^{64}Zn$ | 0.080 | 4.1 |
| | | $^{65}Zn$ | … | 3.0 |
| | | $^{66}Ge$ | … | 3.0 |
| | | $^{67}Ga$ | … | 3.7 |
| | | $^{73}Se$ | … | 0.49 |
| | | $^{74}Se$ | … | 0.50 |
| | | $^{77}Kr$ | … | 0.42 |
| | | $^{78}Kr$ | … | 0.42 |
| | | $^{79}Kr$ | … | 0.42 |
| | | $^{80}Sr$ | … | 0.47 |
| | | $^{81}Rb$ | … | 0.40 |
| | | $^{82}Sr$ | … | 0.41 |
| | | $^{83}Sr$ | … | 0.42 |
| | | $^{84}Sr$ | … | 0.42 |
| | | $^{85}Y$ | … | 0.42 |
| | | $^{86}Zr$ | … | 0.40 |
| | | $^{87}Zr$ | … | 0.39 |
| | | $^{88}Zr$ | … | 0.38 |
| | | $^{89}Nb$ | … | 0.38 |
| | | $^{90}Mo$ | … | 0.37 |
| | | $^{91}Nb$ | … | 0.36 |
| | | $^{92}Mo$ | … | 0.36 |

| | | $^{93}$Tc | ... | 0.36 |
| | | $^{94}$Mo | ... | 0.35 |
| | | $^{95}$Ru | ... | 0.34 |
| | | $^{96}$Ru | ... | 0.33 |
| | | $^{97}$Ru | ... | 0.31 |
| | | $^{98}$Ru | ... | 0.30 |
| | | $^{99}$Rh | ... | 0.29 |
| | | $^{100}$Pd | ... | 0.30 |
| | | $^{101}$Pd | ... | 0.29 |
| | | $^{102}$Pd | ... | 0.29 |
| | | $^{103}$Ag | ... | 0.30 |
| | | $^{104}$Ag | ... | 0.32 |
| Fo8 | $^{45}$Cr(p,γ)$^{46}$Mn | $^{45}$Ti | 0.50 | ... |
| So1 | $^{42}$Ti(p,γ)$^{43}$V | $^{42}$Ca | ... | 2.1 |
| | $^{64}$Ge(p,γ)$^{65}$As | $^{64}$Zn | 0.35 | ... |
| | $^{98}$Cd(p,γ)$^{99}$In | $^{98}$Ru | ... | 6.5 |
| | $^{106}$Sn(p,γ)$^{107}$Sb | $^{107}$Cd | 3.1 | ... |
| | | $^{108}$Cd | 3.0 | ... |
| hiT | $^{30}$S(p,γ)$^{31}$Cl | $^{30}$Si | 0.49 | ... |
| | $^{46}$Cr(p,γ)$^{47}$Mn | $^{46}$Ti | ... | 3.6 |
| | $^{60}$Zn(p,γ)$^{61}$Ga | $^{60}$Ni | 0.42 | 2.2 |
| | $^{64}$Ge(p,γ)$^{65}$As | $^{64}$Zn | 0.05 | 2.9 |
| | | $^{67}$Ga | ... | 2.3 |
| | $^{68}$Se(p,γ)$^{69}$Br | $^{68}$Ge | 0.37 | ... |
| | | $^{69}$Ge | 0.39 | ... |
| | | $^{70}$Ge | 0.39 | ... |
| | | $^{71}$As | 0.37 | ... |
| | $^{105}$Sn(p,γ)$^{106}$Sb | $^{106}$Cd | ... | 0.49 |
| lowT | $^{42}$Ti(p,γ)$^{43}$V | $^{42}$Ca | ... | 3.1 |
| | $^{46}$Cr(p,γ)$^{47}$Mn | $^{46}$Ti | 0.46 | ... |
| | $^{55}$Ni(p,γ)$^{56}$Cu | $^{55}$Co | 0.27 | 2.2 |
| | $^{60}$Zn(p,γ)$^{61}$Ga | $^{60}$Ni | 0.41 | 2.4 |
| | $^{64}$Ge(p,γ)$^{65}$As | $^{64}$Zn | 0.18 | ... |
| | | $^{73}$Se | 2.7 | ... |
| | | $^{74}$Se | 2.7 | ... |
| | | $^{75}$Br | 2.6 | ... |
| | | $^{76}$Kr | 2.3 | ... |
| | | $^{77}$Kr | 3.1 | ... |
| | | $^{78}$Kr | 3.1 | ... |
| | | $^{79}$Kr | 3.1 | ... |
| | | $^{80}$Sr | 2.7 | ... |
| | | $^{81}$Rb | 3.2 | ... |
| | | $^{82}$Sr | 3.2 | ... |
| | | $^{83}$Sr | 3.2 | ... |
| | | $^{84}$Sr | 3.2 | ... |
| | | $^{85}$Y | 3.2 | ... |
| | | $^{86}$Zr | 3.4 | ... |
| | | $^{87}$Zr | 3.5 | ... |
| | | $^{88}$Zr | 3.7 | ... |
| | | $^{89}$Nb | 3.6 | ... |
| | | $^{90}$Mo | 3.7 | ... |
| | | $^{91}$Nb | 3.8 | ... |
| | | $^{92}$Mo | 3.7 | ... |
| | | $^{93}$Tc | 3.7 | ... |
| | | $^{94}$Mo | 3.6 | ... |
| long | $^{64}$Ge(p,γ)$^{65}$As | $^{64}$Zn | 0.21 | 7.1 |
| | | $^{65}$Zn | 0.27 | 2.3 |
| | | $^{66}$Ge | 0.23 | 4.8 |
| | | $^{67}$Ga | ... | 15 |
| | | $^{69}$Ge | ... | 0.47 |
| | | $^{71}$As | ... | 0.34 |
| | | $^{73}$As | ... | 0.38 |
| | | $^{73}$Se | ... | 0.40 |
| | | $^{77}$Br | ... | 0.47 |
| | | $^{77}$Kr | ... | 0.49 |
| | | $^{81}$Kr | ... | 0.46 |
| | | $^{81}$Rb | ... | 0.49 |
| | | $^{84}$Sr | ... | 0.25 |
| | | $^{88}$Zr | ... | 0.48 |
| | $^{89}$Ru(p,γ)$^{90}$Rh | $^{89}$Zr | 0.42 | ... |
| | | $^{89}$Nb | 0.41 | ... |
| short | $^{25}$Si(p,γ)$^{26}$P | $^{25}$Mg | 0.31 | ... |
| | | $^{27}$Al | 2.2 | ... |
| | $^{26}$P(p,γ)$^{27}$S | $^{25}$Mg | 0.11 | ... |
| | | $^{27}$Al | 2.6 | ... |
| | $^{30}$S(p,γ)$^{31}$Cl | $^{30}$Si | 0.48 | ... |
| | $^{42}$Ti(p,γ)$^{43}$V | $^{42}$Ca | 0.15 | 21 |
| | | $^{44}$Ti | ... | 0.49 |
| | $^{46}$Cr(p,γ)$^{47}$Mn | $^{46}$Ti | 0.13 | 2.6 |
| | | $^{48}$Cr | ... | 0.18 |
| | $^{50}$Fe(p,γ)$^{51}$Co | $^{51}$Cr | 4.9 | ... |
| | | $^{52}$Fe | 2.0 | ... |
| | $^{55}$Ni(p,γ)$^{56}$Cu | $^{55}$Co | 0.33 | 6.5 |
| | $^{64}$Ge(p,γ)$^{65}$As | $^{64}$Zn | 0.29 | ... |
| | | $^{65}$Zn | 9.6 | ... |
| | | $^{66}$Ge | 7.8 | ... |
| | | $^{67}$Ga | 5.7 | ... |
| | | $^{68}$Ge | 10 | 0.48 |
| | | $^{69}$Ge | 10 | 0.48 |
| | | $^{70}$Ge | 10 | 0.48 |
| | | $^{71}$As | 11 | 0.46 |
| | | $^{72}$Se | 11 | 0.32 |

| | | $^{76}$Kr | 11 | 0.24 |
|---|---|---|---|---|
| lowZ | $^{26}$P(p,γ)$^{27}$S | $^{25}$Mg | 0.36 | ... |
| | $^{42}$Ti(p,γ)$^{43}$V | $^{42}$Ca | ... | 7.4 |
| | $^{46}$Cr(p,γ)$^{47}$Mn | $^{46}$Ti | 0.24 | ... |
| | $^{55}$Ni(p,γ)$^{56}$Cu | $^{55}$Co | ... | 3.8 |
| | $^{60}$Zn(p,γ)$^{61}$Ga | $^{60}$Ni | 0.48 | ... |
| | $^{64}$Ge(p,γ)$^{65}$As | $^{64}$Zn | 0.080 | 4.0 |
| | | $^{65}$Zn | ... | 2.9 |
| | | $^{66}$Ge | ... | 2.9 |
| | | $^{67}$Ga | ... | 3.5 |
| | | $^{73}$Se | ... | 0.49 |
| | | $^{77}$Kr | ... | 0.42 |
| | | $^{78}$Kr | ... | 0.43 |
| | | $^{79}$Kr | ... | 0.43 |
| | | $^{80}$Sr | ... | 0.48 |
| | | $^{81}$Rb | ... | 0.4 |
| | | $^{82}$Sr | ... | 0.41 |
| | | $^{83}$Sr | ... | 0.42 |
| | | $^{84}$Sr | ... | 0.42 |
| | | $^{85}$Y | ... | 0.43 |
| | | $^{86}$Zr | ... | 0.41 |
| | | $^{87}$Zr | ... | 0.40 |
| | | $^{88}$Zr | ... | 0.39 |
| | | $^{89}$Nb | ... | 0.38 |
| | | $^{90}$Mo | ... | 0.37 |
| | | $^{91}$Nb | ... | 0.36 |
| | | $^{92}$Mo | ... | 0.36 |
| | | $^{93}$Tc | ... | 0.35 |
| | | $^{94}$Mo | ... | 0.34 |
| | | $^{95}$Ru | ... | 0.34 |
| | | $^{96}$Ru | ... | 0.33 |
| | | $^{97}$Ru | ... | 0.31 |
| | | $^{98}$Ru | ... | 0.30 |
| | | $^{99}$Rh | ... | 0.29 |
| | | $^{100}$Pd | ... | 0.30 |
| | | $^{101}$Pd | ... | 0.29 |
| | | $^{102}$Pd | ... | 0.30 |
| | | $^{103}$Ag | ... | 0.31 |
| | | $^{104}$Ag | ... | 0.34 |
| hiZ | $^{64}$Ge(p,γ)$^{65}$As | $^{64}$Zn | 0.24 | ... |
| | | $^{66}$Ge | 0.32 | ... |
| | | $^{68}$Ge | 2.1 | ... |
| | | $^{69}$Ge | 2.1 | ... |
| | | $^{70}$Ge | 2.1 | ... |
| hiZ2 | $^{64}$Ge(p,γ)$^{65}$As | $^{68}$Ge | 3.1 | ... |
| | | $^{69}$Ge | 3.1 | ... |
| | | $^{70}$Ge | 3.2 | ... |
| | | $^{71}$As | 3.1 | ... |
| | | $^{72}$Se | 4.2 | ... |
| | | $^{73}$Se | 4.0 | ... |
| | | $^{76}$Kr | 4.5 | ... |
| | | $^{80}$Sr | 4.6 | ... |

Table IV: Summary of reactions whose ΔQ significantly affect XRB nucleosynthesis in our models. These are the only reactions with Q < 1 MeV that modify the final XRB yield of at least one isotope by at least a factor of two in at least one model, when their nominal Q-values are varied by ±ΔQ. ΔQ for the $^{64}$Ge(p,γ)$^{65}$As reaction affects by far the most final XRB yields (see Table III) in the most models. All Q-values and ΔQ are from [54]; only Q ($^{30}$S(p,γ)$^{31}$Cl) and Q ($^{60}$Zn(p,γ)$^{61}$Ga) are experimental (the others have been estimated from systematic trends).

| Reaction | Q ± ΔQ (keV) | Model Affected |
|---|---|---|
| $^{25}$Si(p,γ)$^{26}$P | 140 ± 196 | Short |
| $^{26}$P(p,γ)$^{27}$S | 719 ± 281 | K04, lowZ[a], short |
| $^{30}$S(p,γ)$^{31}$Cl | 294 ± 50 | hiT, short |
| $^{42}$Ti(p,γ)$^{43}$V | 192 ± 233 | S01, lowT, lowZ, short |
| $^{45}$Cr(p,γ)$^{46}$Mn | 694 ± 515 | F08 |
| $^{46}$Cr(p,γ)$^{47}$Mn | 78 ± 160 | K04, lowT, hiT, lowZ, short |
| $^{50}$Fe(p,γ)$^{51}$Co | 88 ± 161 | Short |
| $^{55}$Ni(p,γ)$^{56}$Cu | 555 ± 140 | K04, lowT, lowZ, short |
| $^{60}$Zn(p,γ)$^{61}$Ga | 192 ± 54 | K04, lowT, hiT[a], lowZ |
| $^{64}$Ge(p,γ)$^{65}$As | −80 ± 300 | K04[a], S01[a], lowT[a], hiT[a], lowZ[a], hiZ, hiZ2, long[a], short |
| $^{68}$Se(p,γ)$^{69}$Br | −450 ± 100 | hiT |
| $^{89}$Ru(p,γ)$^{90}$Rh | 992 ± 711 | Long |
| $^{98}$Cd(p,γ)$^{99}$In | 932 ± 408 | S01 |
| $^{105}$Sn(p,γ)$^{106}$Sb | 357 ± 323 | hiT |
| $^{106}$Sn(p,γ)$^{107}$Sb | 518 ± 302 | S01[a] |

[a] variation of this reaction Q-value affects the nuclear energy generation rate in this model (see text)

FIGURE CAPTIONS

FIG. 1: Final abundances (as mass fractions $X_{f,std}$) versus mass number A, as calculated with the 'K04', 'F08' and 'S01' XRB models (see Tables I and II). These final yields were obtained using standard rates in our nuclear network.

FIG. 2: Final abundances (as mass fractions $X_{f,std}$) versus mass number A, as calculated with the 'long' and 'short' XRB models (see Tables I and II). The yield distribution from the 'K04' model is also shown here as a reference. These final yields were obtained using standard rates in our nuclear network.

FIG. 3: Final abundances (as mass fractions $X_{f,std}$) versus mass number A, as calculated with the 'hiT' and 'lowT' XRB models (see Tables I and II). The yield distribution from the 'K04' model is also shown here as a reference. These final yields were obtained using standard rates in our nuclear network.

FIG. 4: Final abundances (as mass fractions $X_{f,std}$) versus mass number A, as calculated with the 'lowZ', 'hiZ' and 'hiZ2' XRB models (see Tables I and II). Note that the yield distribution from the 'lowZ' model is very similar to that from the 'K04' model. These final yields were obtained using standard rates in our nuclear network.

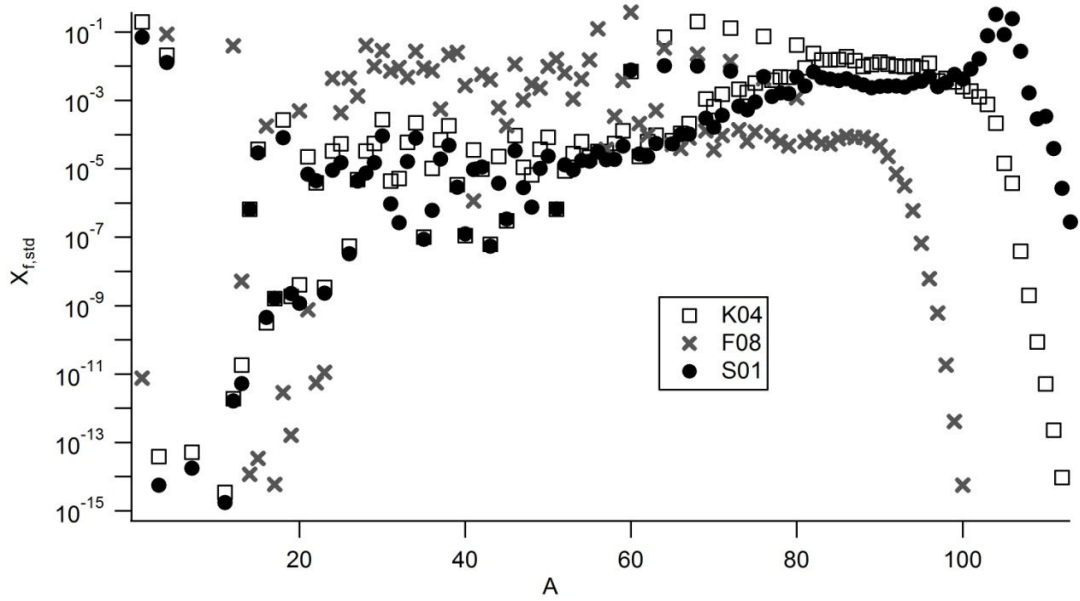

FIGURE 1, A. PARIKH, PHYSICAL REVIEW C

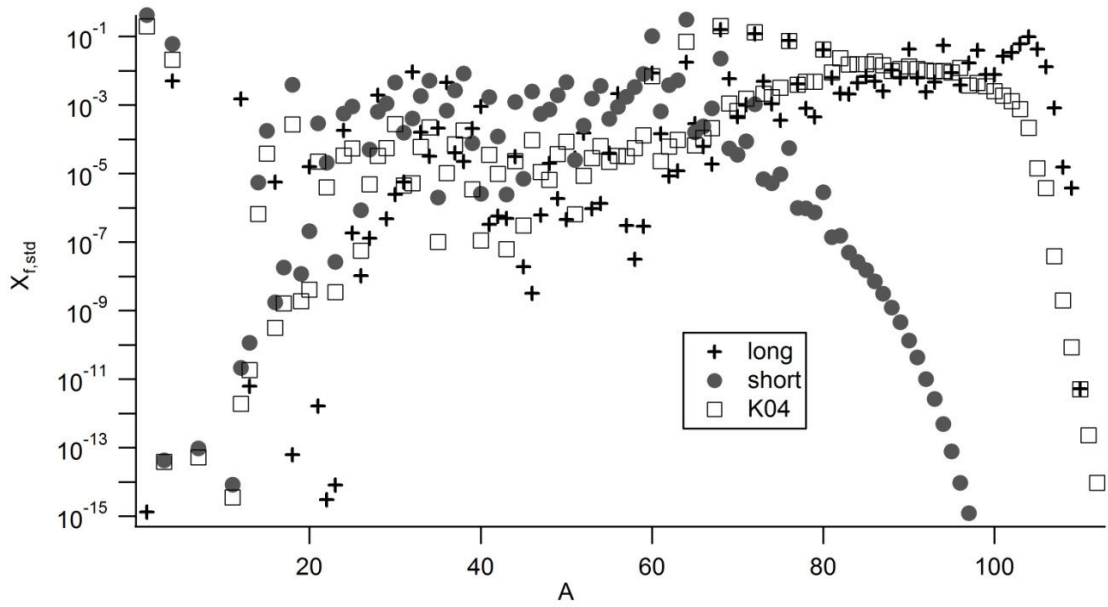

FIGURE 2, A. PARIKH, PHYSICAL REVIEW C

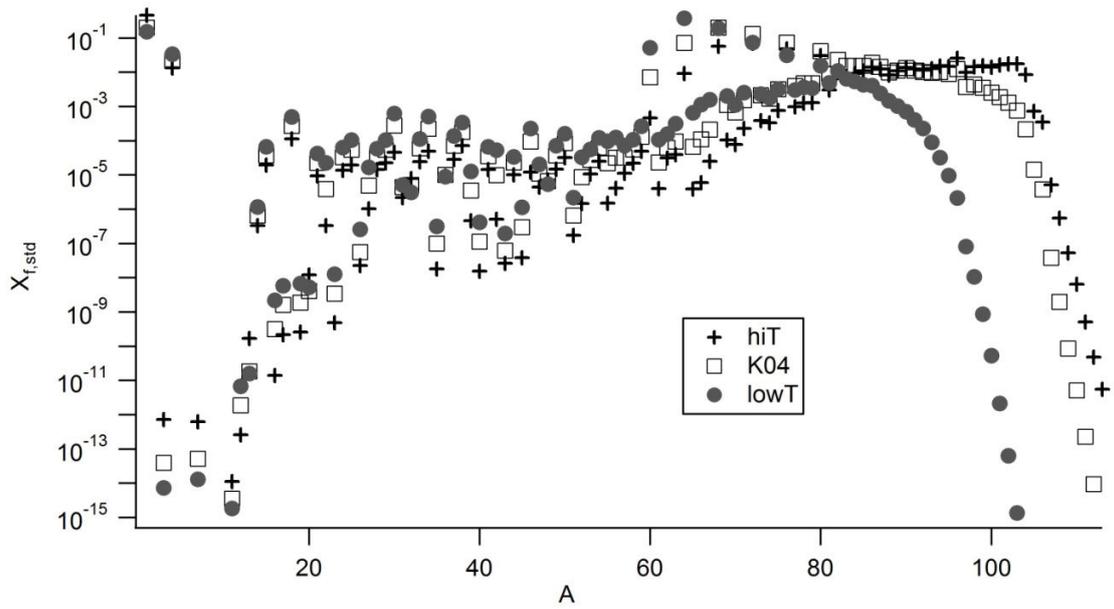

FIGURE 3, A. PARIKH, PHYSICAL REVIEW C

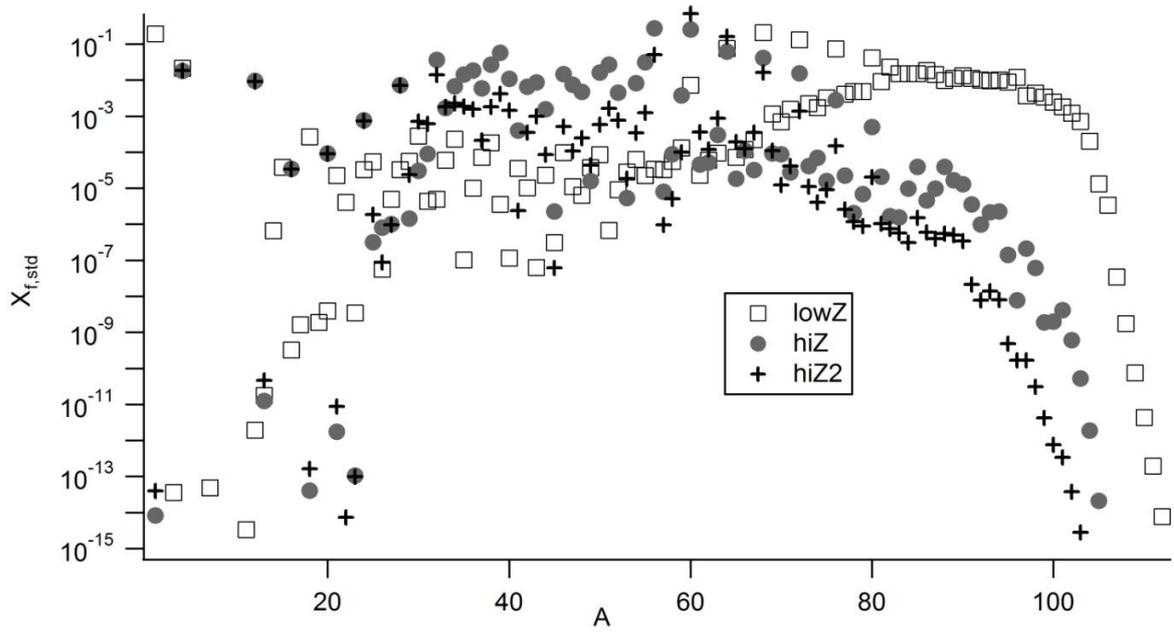

FIGURE 4, A. PARIKH, PHYSICAL REVIEW C

31